\documentclass[aps,prd,twocolumn,nofootinbib,superscriptaddress,floatfix]{revtex4-2}

\usepackage[T1]{fontenc}
\usepackage[utf8]{inputenc}
\usepackage{amsmath,amssymb,amsfonts}
\usepackage{graphicx}
\usepackage{booktabs}
\usepackage{hyperref}
\usepackage{bm}
\usepackage{mathtools}
\usepackage{xcolor}
\usepackage{placeins}
\usepackage{orcidlink}
\usepackage[nameinlink,noabbrev]{cleveref}

\hypersetup{
  colorlinks=true,
  linkcolor=blue,
  citecolor=blue,
  urlcolor=blue
}

\newcommand{\FigPlaceholder}[2]{%
\IfFileExists{#1}{%
  \includegraphics[width=\linewidth]{#1}%
}{%
  \fbox{%
    \parbox[c][4.5cm][c]{0.92\linewidth}{%
      \centering
      Placeholder for #2\\[4pt]
      File \texttt{#1} not found
    }%
  }%
}%
}

\newcommand{\dd}{\mathrm{d}}
\newcommand{\ee}{\mathrm{e}}
\newcommand{\GB}{\mathcal{G}}
\newcommand{\Ord}{\mathcal{O}}
\newcommand{\rh}{r_{\mathrm H}}
\newcommand{\ac}{\alpha_{\mathrm c}}
\newcommand{\Minf}{M_{\mathrm{ADM}}}

\begin{document}

\title{Weakly nonlinear strong-field signatures of spontaneously vectorized Einstein--vector--Gauss--Bonnet black holes near the Schwarzschild bifurcation}

\author{Sardor~Murodov\orcidlink{0000-0003-2360-4475}}
\email{s.murodov@newuu.uz}
\affiliation{New Uzbekistan University, Movarounnahr Street 1, Tashkent 100000, Uzbekistan}
\affiliation{Tashkent State Technical University, Tashkent 100095, Uzbekistan}

\author{Bekzod~Rahmatov\orcidlink{0009-0001-0394-650X}}
\email{rahmatovbekzod@samdu.uz}
\affiliation{University of Tashkent for Applied Sciences, Str. Gavhar 1, Tashkent 100149, Uzbekistan}

\author{Islom~Egamberdiev\orcidlink{0009-0001-6156-9271}}
\email{egamberdiyev.islom@samdaqu.edu.uz}
\affiliation{Samarkand State Technical University named after Mirzo Ulugbek, Lolazor street 70, Samarqand 140143, Uzbekistan}

\author{Dilnavoz~Kamalova\orcidlink{0009-0001-0004-3372}}
\email{kamalovadilnavoz77@gmail.com}
\affiliation{Navoi state university, Ibn Sino street 45, Navoi 210100, Uzbekistan}

\author{Javlon~Rayimbaev\orcidlink{0000-0001-9293-1838}}
\email{javlon@astrin.uz}
\affiliation{Institute of Theoretical Physics, National University of Uzbekistan, Tashkent 100174, Uzbekistan}
\affiliation{Kimyo International University in Tashkent, Shota Rustaveli street 156, Tashkent 100121, Uzbekistan}

\author{Bobomurat~Ahmedov\orcidlink{0000-0002-1232-610X}}
\email{ahmedov@astrin.uz}
\affiliation{Institute of Theoretical Physics, National University of Uzbekistan, Tashkent 100174, Uzbekistan}
\affiliation{School of Physics, Harbin Institute of Technology, Harbin 150001, People’s Republic of China}

\date{\today}

\begin{abstract}
We study the static, spherically symmetric electric branch of spontaneously vectorized Einstein--vector--Gauss--Bonnet black holes close to its Schwarzschild bifurcation. A weakly nonlinear expansion is built around the fundamental vector zero mode. The critical coupling is determined independently by shooting and collocation, the second-order metric correction is checked against the unreduced covariant equations, and the corresponding ADM-mass correction is shown to vanish analytically. The leading departure of the vectorized branch from the bifurcation point follows from a third-order Fredholm solvability condition. Using the resulting metric, we calculate the leading corrections to the photon sphere, shadow scale, marginally bound orbit, innermost stable circular orbit, orbital and epicyclic frequencies, the geometric three-to-two frequency-ratio radius, the instability of the circular photon orbit, and the logarithmic strong-deflection coefficient. Near the bifurcation, the photon sphere and innermost stable orbit move outward, whereas the three-to-two radius moves inward. The photon-orbit angular frequency decreases and its Lyapunov exponent increases. These results characterize the local strong-field geometry of the electric vectorized branch. They are not intended as observational constraints or as an extrapolation to the strongly nonlinear regime.
\end{abstract}

\maketitle

\section{Introduction}\label{sec:introduction}

Strong-field observations provide a direct way to probe the geometry of black holes beyond the weak-field regime in which general relativity (GR) is already tightly constrained. This has motivated both theory-specific calculations and theory-agnostic parametrizations of possible departures from the Schwarzschild and Kerr metrics \citep{Psaltis2008,Will2014,BertiEtAl2015,Krawczynski2012,Krawczynski2018,JohannsenPsaltis2011,Johannsen2013,RezzollaZhidenko2014,VolkelKokkotas2019}. Several of the relevant observables are set by geodesic structure close to the compact object. The instability of the circular photon orbit (ISCO) enters accretion-disk modelling, while unstable photon orbits determine the critical impact parameter that sets the geometric shadow scale \citep{AbdujabbarovRezzollaAhmedov2015,MizunoEtAl2018,EHTM872019,EHTSgrA2022,MurodovRayimbaevAhmedov2023Universe}. Orbital and optical quantities can therefore probe different aspects of the same strong-field geometry.

Spontaneous scalarization is a familiar example of this type of strong-field departure from GR. When a scalar field couples nonminimally to the Gauss--Bonnet invariant, the GR black-hole solution can develop a tachyonic mode; a regular zero mode marks the point where a scalarized branch bifurcates from the GR family \citep{DonevaYazadjiev2018,SilvaEtAl2018,BlazquezSalcedoKleihausKunz2021}. Rotating solutions, massive-scalar extensions, quasinormal modes, stability, and more general coupling functions have been investigated extensively \citep{CunhaEtAl2019,DonevaStaykovYazadjiev2019,DonevaCollodelKrugerEtAl2020,BlazquezSalcedoEtAl2020,AntoniouEtAl2022,KleihausKunzUtermohlen2023,MinamitsujiMukohyamaTsujikawa2024}. For a quadratic coupling, the field perturbation is linear in its amplitude whereas the first metric correction is quadratic, a structure that makes a weakly nonlinear treatment natural near the bifurcation \citep{MotohashiMukohyama2019}.

The corresponding mechanism for a vector field is usually referred to as spontaneous vectorization \citep{Ramazanoglu2017}. Hairy compact objects also arise in a wider class of nonminimally coupled vector--tensor and generalized Proca models, including static and rotating solutions and theories with broken gauge symmetry \citep{Fan2016,Fan2018,BabichevCharmousisHassaine2017,HeisenbergTsujikawa2018,HeisenbergKaseTsujikawa2018,HerdeiroRaduRunarsson2016,RahmanSen2019,ZhouBambiHerdeiro2017,ChannuieMomeni2018}. Related vectorization and tensorization mechanisms have been considered in other settings \citep{RamazanogluUnluturk2019,OliveiraPombo2021,BrihayeHartmannKleihaus2022}. In Einstein--vector--Gauss--Bonnet (EvGB) theory, static spherically symmetric vectorized black holes were constructed by \citet{BartonEtAl2021}, and wormhole solutions are also known for suitable vector--Gauss--Bonnet couplings \citep{BartonKieferKleihaus2022}. The existence and stability of vectorized configurations depend on the detailed field content and coupling structure, as illustrated by related no-go results \citep{Matsumoto2023}. More recently, \citet{KleihausKunz2026} obtained stationary rotating solutions, a static axially symmetric magnetic branch, and the corresponding perturbative bifurcation spectra. For the quadratic, static, spherically symmetric electric branch, the fundamental mode bifurcates from Schwarzschild at
\begin{equation}
 \left(\frac{\lambda}{M^2}\right)_{\mathrm{bif}}
 =4.6817\ldots .
 \label{eq:critical-coupling-intro}
\end{equation}
Earlier work has focused mainly on constructing the solutions and describing their domains of existence, global charges, horizon properties, thermodynamics, and bifurcation spectra \citep{BartonEtAl2021,KleihausKunz2026}. To our knowledge, a single weakly nonlinear calculation starting from the fundamental electric zero mode and carrying the result through to ADM-normalized orbital, timing, and local lensing coefficients has not yet been given. We address that local problem here.

Our calculation starts at the Schwarzschild bifurcation and follows the electric branch to the first nontrivial order in its vector amplitude. We construct the backreacted metric, determine the local direction of the branch from the Fredholm condition, and then evaluate neutral timelike and null geodesic observables on that geometry. The quantities considered include circular-orbit energetics, the characteristic orbital radii, epicyclic frequencies, photon-orbit instability, and the local strong-deflection coefficient. We do not perform an observational fit, nor do we use the perturbative series deep on the nonlinear branch. The static solution is often said to carry an ``electric charge,'' but the EvGB vector is not an ordinary Maxwell gauge field: the $A_\mu A^\mu\GB$ interaction breaks $U(1)$ gauge invariance \citep{KleihausKunz2026}. We therefore consider ordinary matter to be neutral and use metric geodesics throughout.

The EvGB model and the static spherical reduction are summarized in \cref{sec:theory}. Section~\ref{sec:weakly-nonlinear} develops the weakly nonlinear branch, and Secs.~\ref{sec:geodesics} and \ref{sec:shifts} give the geodesic relations and perturbative shift formulas. Numerical details and the resulting coefficients are presented in Secs.~\ref{sec:numerics} and \ref{sec:results}. The scope of the approximation is discussed in Sec.~\ref{sec:discussion}, followed by the conclusions.

\section{Einstein--vector--Gauss--Bonnet theory}\label{sec:theory}

\subsection{Action and field equations}\label{subsec:action}

We work with the EvGB action used in Refs.~\citep{BartonEtAl2021,KleihausKunz2026},
\begin{equation}
 S=\frac{1}{16\pi}\int \dd^4x\,\sqrt{-g}
 \left[
 R-F_{\mu\nu}F^{\mu\nu}
 +\lambda A_\mu A^\mu\GB
 \right],
 \label{eq:action}
\end{equation}
where
\begin{equation}
 F_{\mu\nu}=\nabla_\mu A_\nu-\nabla_\nu A_\mu
 \label{eq:field-strength}
\end{equation}
and
\begin{equation}
 \GB=R_{\mu\nu\rho\sigma}R^{\mu\nu\rho\sigma}
 -4R_{\mu\nu}R^{\mu\nu}+R^2
 \label{eq:GB}
\end{equation}
denote the vector field strength and Gauss--Bonnet invariant. The coupling $\lambda$ has dimensions of length squared. In four dimensions the Gauss--Bonnet term alone is topological, but the factor $A_\mu A^\mu$ makes the interaction dynamical \citep{KleihausKunz2026}.

Since $\lambda$ is the only dimensionful coupling in vacuum, a rescaling of all lengths gives $M\to sM$, $Q\to sQ$, and $\lambda\to s^2\lambda$. It is therefore convenient to label the solution space by
\begin{equation}
 \alpha\equiv\frac{\lambda}{M^2},
 \qquad
 q\equiv\frac{Q}{M}.
 \label{eq:dimensionless-branch-parameters}
\end{equation}
In the perturbative calculation we vary $\alpha$ at fixed reference scale $M_0$. This simply parametrizes the dimensionless branch. If the underlying theory is instead specified by a fixed $\lambda=\lambda_\star$, the same family is obtained by rescaling the solutions, with $M=\sqrt{\lambda_\star/\alpha}$.

Varying the action with respect to the vector field gives, after an integration by parts,
\begin{equation}
 \nabla_\mu F^{\mu\nu}
 =-\frac{\lambda}{2}A^\nu\GB
 \label{eq:vector-eom}
\end{equation}
which agrees with the vector equation reported in Ref.~\citep{KleihausKunz2026}.

The metric variation yields
\begin{equation}
 G_{\mu\nu}=\frac12 T^{\mathrm{eff}}_{\mu\nu},
 \qquad
 T^{\mathrm{eff}}_{\mu\nu}
 =T^{(A)}_{\mu\nu}-2T^{(\mathrm{GB})}_{\mu\nu},
 \label{eq:einstein-eom}
\end{equation}
with the field-strength contribution
\begin{equation}
 T^{(A)}_{\mu\nu}
 =4F_{\mu}{}^{\rho}F_{\nu\rho}
 -g_{\mu\nu}F_{\rho\sigma}F^{\rho\sigma}.
 \label{eq:vector-stress}
\end{equation}
The explicit index form of $T^{(\mathrm{GB})}_{\mu\nu}$ is cumbersome because $A_\rho A^\rho$ itself depends on the inverse metric. We therefore derive the radial equations from the symmetry-reduced action and later check them against the unreduced covariant equations.

\subsection{Static spherical ansatz}\label{subsec:ansatz}

We use the areal radial coordinate $r$ and write
\begin{equation}
\begin{aligned}
 \dd s^2={}&-N(r)\ee^{-2\delta(r)}\dd t^2
 +\frac{\dd r^2}{N(r)}+r^2\dd\Omega^2,\\
 N(r)={}&1-\frac{2m(r)}{r}.
\end{aligned}
\label{eq:metric-ansatz}
\end{equation}
with a purely electric vector profile
\begin{equation}
 A_\mu\dd x^\mu=V(r)\dd t.
 \label{eq:vector-ansatz}
\end{equation}
The determinant, vector norm, and field-strength invariant are
\begin{align}
 \sqrt{-g}&=\ee^{-\delta}r^2\sin\theta,\nonumber\\
 A_\mu A^\mu&=-\frac{\ee^{2\delta}V^2}{N},\nonumber\\
 F_{\mu\nu}F^{\mu\nu}&=-2\ee^{2\delta}(V')^2.
 \label{eq:basic-invariants}
\end{align}
A prime denotes $\dd/\dd r$. The reduced curvature scalars, radial Lagrangian, and generalized Euler--Lagrange operator are given in Appendix~\ref{app:weak-details}; only the equations needed for the perturbative construction are kept in the main text.

\subsection{Exact radial vector equation}\label{subsec:exact-vector}

For the ansatz \cref{eq:metric-ansatz,eq:vector-ansatz}, the $t$ component of \cref{eq:vector-eom} becomes
\begin{equation}
 \frac{1}{\sqrt{-g}}\partial_r
 \left(\sqrt{-g}F^{rt}\right)
 =-\frac{\lambda}{2}A^t\GB.
 \label{eq:vector-step1}
\end{equation}
Using
\begin{equation}
 F^{rt}=-\ee^{2\delta}V',
 \qquad
 A^t=-\frac{\ee^{2\delta}}{N}V,
 \label{eq:vector-components}
\end{equation}
we obtain
\begin{equation}
 V''+\left(\frac{2}{r}+\delta'\right)V'
 +\frac{\lambda\GB}{2N}V=0.
 \label{eq:exact-vector-ode}
\end{equation}
Equation~\eqref{eq:exact-vector-ode} is used both for the linear zero mode and, at third order, for the solvability condition.

\section{Weakly nonlinear construction near the bifurcation}\label{sec:weakly-nonlinear}

\subsection{Schwarzschild branch and perturbative bookkeeping}\label{subsec:bookkeeping}

At the bifurcation the background solution is Schwarzschild,
\begin{equation}
 N_0(r)=f(r)=1-\frac{2M_0}{r},
 \qquad
 \delta_0(r)=0,
 \qquad
 V_0(r)=0,
 \label{eq:schwarzschild-background}
\end{equation}
where $M_0=\rh/2$ is the mass parameter associated with the fixed background horizon $\rh=2M_0$. The Schwarzschild Gauss--Bonnet invariant is
\begin{equation}
 \GB_0=\frac{48M_0^2}{r^6}.
 \label{eq:GB-schwarzschild}
\end{equation}

We define a dimensionless vector amplitude $\varepsilon$ through the asymptotic charge and expand
\begin{align}
 V(r)&=\varepsilon v_1(r)+\varepsilon^3v_3(r)+\Ord(\varepsilon^5),
 \label{eq:V-expansion}\\
 m(r)&=M_0+\varepsilon^2m_2(r)+\Ord(\varepsilon^4),
 \label{eq:m-expansion}\\
 \delta(r)&=\varepsilon^2\delta_2(r)+\Ord(\varepsilon^4),
 \label{eq:delta-expansion}\\
 \frac{\lambda}{M_0^2}
 &=\ac+\varepsilon^2\alpha_2^{(\mathrm H)}
 +\Ord(\varepsilon^4).
 \label{eq:alpha-expansion}
\end{align}
The superscript $(\mathrm H)$ denotes the fixed-horizon scheme used at this stage. The expansion follows the discrete symmetry $A_\mu\rightarrow-A_\mu$: the vector field is odd in $\varepsilon$, while the metric and coupling correction are even.

We normalize the fundamental mode according to
\begin{equation}
 v_1(r)=\frac{M_0}{r}+\Ord(r^{-5})
 \qquad (r\rightarrow\infty),
 \label{eq:v-normalization}
\end{equation}
so that
\begin{equation}
 Q=\varepsilon M_0+\Ord(\varepsilon^3).
 \label{eq:charge-amplitude}
\end{equation}
Here $Q$ is the asymptotic EvGB vector charge, not an ordinary Maxwell charge.

\subsection{First-order vector zero mode}\label{subsec:first-order}

Substituting \cref{eq:schwarzschild-background,eq:V-expansion,eq:alpha-expansion} into \cref{eq:exact-vector-ode} and retaining $\Ord(\varepsilon)$ terms gives
\begin{equation}
 v_1''+\frac{2}{r}v_1'
 +\frac{24\lambda_{\mathrm c}M_0^2}{r^6f}v_1=0,
 \qquad
 \lambda_{\mathrm c}=\ac M_0^2.
 \label{eq:zero-mode-ode}
\end{equation}
Equivalently,
\begin{equation}
 \mathcal L_{\mathrm c}[v_1]
 \equiv
 \frac{1}{r^2}\left(r^2v_1'\right)'
 +\frac{\lambda_{\mathrm c}\GB_0}{2f}v_1=0.
 \label{eq:self-adjoint-operator}
\end{equation}
Regularity of the physical vector norm at the future horizon and asymptotic flatness imply
\begin{equation}
 v_1(2M_0)=0,
 \qquad
 v_1(\infty)=0.
 \label{eq:zero-mode-bcs}
\end{equation}
The normalization is fixed by \cref{eq:v-normalization}. Nontrivial solutions occur only at discrete values of $\lambda/M_0^2$; for the fundamental nodeless mode \citep{KleihausKunz2026},
\begin{equation}
 \ac=4.6817469\ldots\, .
 \label{eq:critical-coupling}
\end{equation}
The areal-coordinate problem \cref{eq:zero-mode-ode,eq:zero-mode-bcs} is equivalent to the isotropic-coordinate formulation of Ref.~\citep{KleihausKunz2026}. The Schwarzschild coordinates are related by
\begin{equation}
 r=\rho\left(1+\frac{M_0}{2\rho}\right)^2,
 \qquad
 \rho_{\mathrm H}=\frac{M_0}{2}.
 \label{eq:isotropic-transform}
\end{equation}

Near the horizon the regular solution is analytic in $r-2M_0$. The coefficients used to initialize the shooting integration are listed in Appendix~\ref{app:weak-details}.

\subsection{Second-order metric backreaction}\label{subsec:second-order}

At $\Ord(\varepsilon^2)$ the zero mode backreacts on the metric. The reduced field equations become two first-order equations for $m_2$ and $\delta_2$. Their source terms, given in Appendix~\ref{app:weak-details}, contain $v_1$ and its derivatives. Apparent horizon singularities cancel once the regular zero-mode expansion is imposed.

We work first in a horizon-fixed scheme,
\begin{equation}
 m_2(2M_0)=0,
 \qquad
 \delta_2(\infty)=0.
 \label{eq:second-order-bcs}
\end{equation}
The first condition keeps the horizon at $r=2M_0$; the second fixes the normalization of the asymptotic time coordinate. The corresponding finite horizon derivatives are listed in Appendix~\ref{app:weak-details}.
At infinity,
\begin{align}
 m_2(r)&=\mu_2-\frac{M_0^2}{2r}
 +\frac{8M_0^3\lambda_{\mathrm c}}{r^4}
 +\Ord(r^{-5}),
 \label{eq:m2-asymptotic}\\
 \delta_2(r)&=-\frac{24M_0^3\lambda_{\mathrm c}}{5r^5}
 +\Ord(r^{-6}),
 \label{eq:delta2-asymptotic}
\end{align}
where
\begin{equation}
 \mu_2\equiv m_2(\infty)
 \label{eq:mu2}
\end{equation}
is the second-order ADM-mass correction in the fixed-horizon scheme. It vanishes identically. The zero-mode equation turns the mass equation into a total radial derivative whose boundary term is zero both at the regular horizon and at infinity (Appendix~\ref{app:weak-details}), so
\begin{equation}
 \mu_2=0
 \qquad \text{at }\Ord(\varepsilon^2).
 \label{eq:mu2-exact}
\end{equation}
This statement applies only to the $\Ord(\varepsilon^2)$ coefficient of the local expansion; it does not imply a constant ADM mass along the full nonlinear branch.

The metric relevant for geodesics is
\begin{equation}
 A(r)\equiv N\ee^{-2\delta}
 =f(r)+\varepsilon^2a(r)+\Ord(\varepsilon^4),
 \label{eq:A-expansion}
\end{equation}
with
\begin{equation}
 a(r)=-\frac{2m_2(r)}{r}-2f(r)\delta_2(r).
 \label{eq:a-metric-correction}
\end{equation}
The radial metric function is
\begin{equation}
 N(r)=f(r)+\varepsilon^2 n(r)+\Ord(\varepsilon^4),
 \qquad
 n(r)=-\frac{2m_2(r)}{r}.
 \label{eq:n-correction}
\end{equation}

\subsection{Third-order solvability condition}\label{subsec:third-order}

The distance from the bifurcation is fixed by the third-order vector equation. At $\Ord(\varepsilon^3)$ this equation contains the second-order metric correction and the $\Ord(\varepsilon^2)$ shift of the coupling. Since the zero-mode operator is self-adjoint, regular homogeneous boundary conditions for $v_3$ lead to a Fredholm solvability condition. This determines the quadratic coefficient governing the local direction of the vectorized branch. The full source and integral condition are given in Appendix~\ref{app:weak-details}; their numerical evaluation is described in Sec.~\ref{sec:numerics}.

\subsection{ADM normalization and fixed-coupling interpretation}\label{subsec:ADM-conversion}

In the horizon-fixed scheme,
\begin{equation}
 \Minf=M_0+\varepsilon^2\mu_2+\Ord(\varepsilon^4).
 \label{eq:ADM-mass}
\end{equation}
The dimensionless coupling normalized by the physical ADM mass is consequently
\begin{align}
 \alpha_{\mathrm{ADM}}
 \equiv\frac{\lambda}{\Minf^2}
 &=\ac+\varepsilon^2\widehat\alpha_2
 +\Ord(\varepsilon^4),
 \label{eq:alpha-ADM}\\
 \widehat\alpha_2
 &=\alpha_2^{(\mathrm H)}
 -2\ac\frac{\mu_2}{M_0}.
 \label{eq:alpha2-hat}
\end{align}
Equation~\eqref{eq:mu2-exact} implies
\begin{align}
 \Minf&=M_0+\Ord(\varepsilon^4),\nonumber\\
 \widehat\alpha_2&=\alpha_2^{(\mathrm H)}
 \quad\text{at the retained order}.
 \label{eq:ADM-simplification}
\end{align}
Hence
\begin{equation}
 \varepsilon^2
 =\frac{\alpha_{\mathrm{ADM}}-\ac}{\widehat\alpha_2}
 +\Ord\!\left[(\alpha_{\mathrm{ADM}}-\ac)^2\right].
 \label{eq:epsilon-alpha}
\end{equation}
Because $Q/\Minf=\varepsilon+\Ord(\varepsilon^3)$, the same corrections can be written in terms of $q^2=(Q/\Minf)^2$. Equations~\eqref{eq:alpha-ADM}--\eqref{eq:epsilon-alpha} describe a local family of dimensionless solutions. For fixed physical $\lambda$, increasing $\alpha_{\rm ADM}=\lambda/\Minf^2$ corresponds instead to decreasing $\Minf$; the fundamental coupling itself is not varied.

\section{Circular geodesics and orbital observables}\label{sec:geodesics}

\subsection{Conserved quantities and radial equation}\label{subsec:radial-equation}

Consider the general static spherical line element
\begin{equation}
\begin{aligned}
 \dd s^2={}&-A(r)\dd t^2+B(r)\dd r^2\\
 &+r^2\left(\dd\theta^2+\sin^2\theta\dd\phi^2\right),\\
 B(r)={}&\frac{1}{N(r)}.
\end{aligned}
\label{eq:general-spherical-metric}
\end{equation}
Spherical symmetry allows the orbit to be placed in the equatorial plane, $\theta=\pi/2$. The geodesic Lagrangian is
\begin{equation}
 2\mathcal L=-A\dot t^2+B\dot r^2+r^2\dot\phi^2=-\kappa,
 \label{eq:geodesic-lagrangian}
\end{equation}
where $\kappa=1$ for timelike geodesics and $\kappa=0$ for null geodesics. The Killing symmetries give
\begin{equation}
 E=A\dot t,
 \qquad
 L=r^2\dot\phi.
 \label{eq:EL-conserved}
\end{equation}
Substitution into \cref{eq:geodesic-lagrangian} yields
\begin{equation}
 B\dot r^2=\frac{E^2}{A}-\kappa-\frac{L^2}{r^2}.
 \label{eq:radial-geodesic}
\end{equation}

For timelike motion ($\kappa=1$), multiplying \cref{eq:radial-geodesic} by $A$ gives
\begin{equation}
 AB\dot r^2=E^2-V_{\rm eff}(r),
 \label{eq:radial-effective-potential}
\end{equation}
with the effective potential
\begin{equation}
 V_{\rm eff}(r)=A(r)\left(1+\frac{L^2}{r^2}\right).
 \label{eq:effective-potential}
\end{equation}
At fixed $L$, circular timelike orbits occur at extrema of $V_{\rm eff}$, and radial stability requires a local minimum. We use this form below to display the change of the strong-field potential produced by the EvGB backreaction.

\subsection{Timelike circular orbits}\label{subsec:timelike-circular}

For a circular timelike orbit at $r=r_0$, the radial function and its first derivative vanish:
\begin{equation}
 \frac{E^2}{A}-1-\frac{L^2}{r^2}=0,
 \label{eq:circular1}
\end{equation}
\begin{equation}
 -\frac{E^2A'}{A^2}+\frac{2L^2}{r^3}=0.
 \label{eq:circular2}
\end{equation}
Solving \cref{eq:circular1,eq:circular2} gives
\begin{equation}
 E_c^2=\frac{2A^2}{2A-rA'},
 \qquad
 L_c^2=\frac{r^3A'}{2A-rA'}.
 \label{eq:EL-circular}
\end{equation}
The coordinate angular frequency is
\begin{align}
 \Omega_\phi
 &=\frac{\dot\phi}{\dot t}
 =\frac{LA}{Er^2},
 \nonumber\\
 \Omega_\phi^2=\frac{A'}{2r}.
 \label{eq:Omega-phi}
\end{align}
For Schwarzschild, $A=f$, and \cref{eq:Omega-phi} reduces to $\Omega_{\phi,0}^2=M_0/r^3$. Circular-orbit energetics and the ISCO are central ingredients of relativistic black-hole accretion theory \citep{BardeenPressTeukolsky1972,PageThorne1974,AbramowiczFragile2013,PennaEtAl2010,RemillardMcClintock2006,MurodovEtAl2026KS}.

\subsection{Marginally bound orbit}\label{subsec:mbo}

The marginally bound circular orbit is defined by unit specific energy,
\begin{equation}
 E_c(r_{\rm MBO})=1.
 \label{eq:mbo-energy-condition}
\end{equation}
Using \cref{eq:EL-circular}, this condition is equivalently
\begin{equation}
 \mathcal M(r)\equiv2A^2-2A+rA'=0.
 \label{eq:mbo-condition}
\end{equation}
For Schwarzschild, $r_{\rm MBO}^{(0)}=4M_0$. The marginally bound orbit (MBO) marks the transition between bound ($E_c<1$) and unbound ($E_c>1$) circular motion.

\subsection{Photon sphere and critical impact parameter}\label{subsec:photon-sphere}

For null motion, define the impact parameter $b=L/E$. Equation~\eqref{eq:radial-geodesic} becomes
\begin{equation}
 B\dot r^2=E^2\left(\frac{1}{A}-\frac{b^2}{r^2}\right).
 \label{eq:null-radial}
\end{equation}
A circular photon orbit obeys
\begin{equation}
 \frac{1}{A}-\frac{b^2}{r^2}=0,
 \qquad
 -\frac{A'}{A^2}+\frac{2b^2}{r^3}=0.
 \label{eq:null-circular-conditions}
\end{equation}
Eliminating $b$ gives
\begin{equation}
 \mathcal P(r)\equiv rA'-2A=0,
 \label{eq:photon-condition}
\end{equation}
while the critical impact parameter is
\begin{equation}
 b_{\mathrm{ph}}^2=\frac{r_{\mathrm{ph}}^2}{A(r_{\mathrm{ph}})}.
 \label{eq:critical-impact}
\end{equation}
For a static spherical black hole, this critical impact parameter sets the geometric shadow boundary. The relation between unstable null orbits and observable shadow structure has been developed from the classic Schwarzschild analyses to more general photon surfaces and shadow diagnostics \citep{Darwin1959,Synge1966,ClaudelVirbhadraEllis2001,Perlick2004,GalloVillanueva2015,CunhaHerdeiroRadu2017,HiokiMaeda2009,PerlickTsupko2022,GuoLi2020,AhmedMurodovEtAl2026NPB}.

The same circular null orbit defines a coordinate angular frequency
\begin{equation}
 \Omega_{\rm ph}=\frac{\sqrt{A(r_{\rm ph})}}{r_{\rm ph}}=\frac{1}{b_{\rm ph}}.
 \label{eq:omega-photon}
\end{equation}
Its radial instability is characterized by the coordinate-time Lyapunov exponent
\begin{equation}
 \lambda_{\rm L}^2
 =\frac{1}{2B}
 \left[
 -A''+\frac{2(A')^2}{A}-\frac{6A}{r^2}
 \right]_{r=r_{\rm ph}}.
 \label{eq:lyapunov-photon}
\end{equation}
The quantity $\lambda_{\rm L}^{-1}$ is the local coordinate-time instability scale of the circular null orbit \citep{CardosoEtAl2009,KogaHarada2019}. In geometric optics, $\Omega_{\rm ph}$ and $\lambda_{\rm L}$ also enter familiar connections with strong-deflection lensing and eikonal quasinormal-mode estimates \citep{StefanovYazadjiev2010,GuoWangWu2022,Tsupko2022}. In the generic strong-deflection expansion for a static spherical lens \citep{Bozza2002},
\begin{equation}
 \hat\alpha(b)
 =-\bar a\ln\!\left(\frac{b}{b_{\rm ph}}-1\right)+\bar b+\cdots,
 \label{eq:strong-deflection}
\end{equation}
the logarithmic slope can be written in the present notation as
\begin{equation}
 \bar a=\frac{\Omega_{\rm ph}}{\lambda_{\rm L}}.
 \label{eq:abar-omega-lambda}
\end{equation}
We use only $\bar a$. The regular coefficient $\bar b$ depends on the full deflection integral and cannot be obtained from local photon-sphere data alone.

\subsection{ISCO condition}\label{subsec:isco}

Marginal stability can be imposed by setting the derivative of $L_c^2$ to zero. Differentiating the second expression in \cref{eq:EL-circular} gives
\begin{equation}
 \frac{\dd L_c^2}{\dd r}
 =\frac{2r^2}{(2A-rA')^2}
 \left[3AA'-2r(A')^2+rAA''\right].
 \label{eq:dLdr}
\end{equation}
Hence the ISCO is determined by
\begin{equation}
 \mathcal I(r)\equiv
 3AA'-2r(A')^2+rAA''=0.
 \label{eq:isco-condition}
\end{equation}
For Schwarzschild,
\begin{equation}
 \mathcal I_0(r)=\frac{2M_0(r-6M_0)}{r^3},
 \qquad
 r_{\mathrm{ISCO}}^{(0)}=6M_0.
 \label{eq:isco-schwarzschild}
\end{equation}

The circular-orbit energy and angular momentum at this radius define $E_{\rm ISCO}$ and $L_{\rm ISCO}$. We also use the geodesic binding-efficiency proxy
\begin{equation}
 \eta\equiv1-E_{\rm ISCO}.
 \label{eq:eta-isco}
\end{equation}

\subsection{Radial and vertical epicyclic frequencies}\label{subsec:epicyclic}

Let $r=r_0+\delta r$ and keep $E$ and $L$ fixed at their circular-orbit values. From \cref{eq:radial-geodesic}, the proper-time radial frequency is obtained by expanding the radial function to quadratic order. Converting from proper time to coordinate time with $\dot t=E/A$ yields
\begin{equation}
 \Omega_r^2=\frac{1}{2B}
 \left[
 A''-\frac{2(A')^2}{A}+\frac{3A'}{r}
 \right].
 \label{eq:Omega-r}
\end{equation}
For Schwarzschild,
\begin{equation}
 \Omega_{r,0}^2
 =\frac{M_0}{r^3}\left(1-\frac{6M_0}{r}\right).
 \label{eq:Omega-r-Schwarzschild}
\end{equation}
Spherical symmetry implies
\begin{equation}
 \Omega_\theta=\Omega_\phi,
 \qquad
 \Omega_{\mathrm{nod}}
 =\Omega_\phi-\Omega_\theta=0.
 \label{eq:vertical-frequency}
\end{equation}
The periastron-precession frequency is
\begin{equation}
 \Omega_{\mathrm{per}}=\Omega_\phi-\Omega_r.
 \label{eq:periastron}
\end{equation}

\section{Perturbative strategy for orbital observables}\label{sec:shifts}

We substitute the weakly nonlinear metric into the geodesic relations of Sec.~\ref{sec:geodesics} and truncate every observable consistently at $\Ord(\varepsilon^2)$. If a characteristic radius is defined by $F(r,\varepsilon)=F_0(r)+\varepsilon^2F_2(r)+\cdots$, its leading displacement is
\begin{equation}
 \delta r=-\frac{F_2(r_0)}{F_0'(r_0)}.
 \label{eq:root-shift}
\end{equation}
We use Eq.~\eqref{eq:root-shift} for the photon sphere, MBO, and ISCO, and then evaluate impact parameters and frequencies at the shifted radii. The explicit second-order formulas are collected in Appendix~\ref{app:orbital-derivations}.

For a static spherical metric $\Omega_\theta=\Omega_\phi$. Ratios of orbital and epicyclic frequencies, especially $3{:}2$, appear in many high-frequency Quasi-Periodic Oscillation (QPO) models \citep{AbramowiczKluzniak2001,KluzniakAbramowicz2002,StellaVietri1998,StellaVietriMorsink1999,TorokEtAl2011,SramkovaEtAl2015,StuchlikKolos2015,StuchlikKolos2016,Stefanov2014,MottaEtAl2014,MurodovEtAl2026EsGBQPO}. We use the $3{:}2$ condition only to define a geometric radius and do not assume any disk-resonance mechanism:
\begin{equation}
 \frac{\Omega_\phi}{\Omega_r}=\frac32,
 \qquad
 \mathcal H(r)\equiv4\Omega_\phi^2-9\Omega_r^2=0,
 \label{eq:resonance-condition}
\end{equation}
For Schwarzschild this gives $r_{3:2}^{(0)}=54M_0/5$. Appendix~\ref{app:orbital-derivations} gives its second-order shift and the ADM normalization of the resulting observables. We do not attempt a QPO fit at this stage; such an application should first be supported by a comparison with the full numerical EvGB branch.

\section{Numerical implementation}\label{sec:numerics}

We set $M_0=1$ and use the compact coordinate
\begin{equation}
 x=1-\frac{2M_0}{r}\in[0,1].
 \label{eq:compact-coordinate}
\end{equation}
In this coordinate the zero-mode equation becomes
\begin{equation}
 \frac{\dd^2v_1}{\dd x^2}
 +\frac{3\alpha}{2}\frac{(1-x)^2}{x}v_1=0,
 \label{eq:zero-mode-compact}
\end{equation}
with $v_1(0)=v_1(1)=0$ and $v_{1,x}(1)=-1/2$. We solved this eigenvalue problem independently with an adaptive shooting method and with a collocation boundary-value solver in which $\alpha$ is treated as an eigenparameter. The two calculations give
\begin{equation}
 \alpha_{\rm c}^{\rm shoot}=4.681746973049,
 \qquad
 \alpha_{\rm c}^{\rm BVP}=4.681746973049,
 \label{eq:alpha-numerical-check}
\end{equation}
and agree to the precision shown, as well as with the published perturbative spectrum \citep{KleihausKunz2026}.

We integrated the second-order equations in the same compact coordinate. The mass function was evolved outward from $m_2(0)=0$, while $\delta_2$ was integrated inward from $\delta_2(1)=0$ to maintain accuracy in the asymptotic tail. The numerical solution reproduces the expansions in \cref{eq:m2-asymptotic,eq:delta2-asymptotic}. As a check of the exact identity \eqref{eq:mu2-exact}, we find
\begin{equation}
 \frac{m_2(\infty)}{M_0}=-3.7\times10^{-14},
 \label{eq:mu2-numerical}
\end{equation}
which is consistent with zero at the level of numerical roundoff and integration error.

For the Fredholm integral, derivatives entering $\GB_2$ were evaluated analytically from the compact-coordinate system rather than by finite differences. Adaptive quadrature of \cref{eq:alpha2-solvability} gives
\begin{equation}
 \alpha_2^{(\mathrm H)}=11.060673,
 \qquad
 \widehat\alpha_2=11.060673,
 \label{eq:alpha2-numerical}
\end{equation}
The equality follows from \cref{eq:mu2-exact,eq:alpha2-hat}. The scripts and figure data are included with the source material.

We varied the horizon cutoff over $10^{-6}\le x_0\le10^{-4}$; the shooting and collocation eigenvalues remain stable at roughly the $10^{-13}$ level. The full calculation was then repeated while changing the shooting and second-order cutoffs, integration and quadrature tolerances, and the maximum integration step. Table~\ref{tab:convergence-audit} summarizes the resulting changes. Even for the deliberately coarse choice $x_0=10^{-6}$, the variation remains below $1.7\times10^{-8}$ in $\widehat\alpha_2$ and below $4\times10^{-9}$ in the quoted orbital coefficients. We also checked the second-order radial equations against independent components of the unreduced covariant metric equations; the symbolic identities are given in Appendix~\ref{subsec:full-equation-check}.

\begin{table}[t]
\centering
\caption{Numerical convergence audit. The last column is the maximum absolute change from the baseline calculation across variations of the horizon cutoffs, integration tolerances, quadrature tolerance, and maximum step size described in the text.}
\label{tab:convergence-audit}
\begin{tabular}{lcc}
\toprule
Quantity & Baseline & Max. absolute change\\
\midrule
$\alpha_c$ & $4.681746973049$ & $2.4\times10^{-14}$\\
$\widehat\alpha_2$ & $11.060673$ & $1.64\times10^{-8}$\\
$C_q(r_{\rm ph}/M)$ & $1.78834935$ & $1.06\times10^{-9}$\\
$C_q(b_{\rm ph}/M)$ & $0.90589866$ & $1.83\times10^{-9}$\\
$C_q(r_{\rm ISCO}/M)$ & $0.38726014$ & $2.11\times10^{-9}$\\
$S_q(M\Omega_{\rm ISCO})$ & $-0.137423046$ & $3.51\times10^{-10}$\\
$C_q(r_{3:2}/M)$ & $-1.79115679$ & $3.80\times10^{-9}$\\
\bottomrule
\end{tabular}
\end{table}

\section{Numerical results}\label{sec:results}

Since $Q/M=\varepsilon+\mathcal O(\varepsilon^3)$ and $\mu_2=0$ at second order, we express the final results in terms of $q=Q/M$. The Fredholm condition gives
\begin{equation}
 \frac{\lambda}{M^2}
 =4.681746973049+11.060673\,q^2+\mathcal O(q^4).
 \label{eq:branch-relation-final}
\end{equation}
The positive coefficient means that the electric branch initially extends toward larger $\lambda/M^2$, consistent with the nonlinear solutions of Refs.~\citep{BartonEtAl2021,KleihausKunz2026}. Here and below $M$ is the ADM mass. If $\lambda$ is kept fixed, the same branch direction corresponds to decreasing $M$ because $M=\sqrt{\lambda/\alpha}$.

The orbital observables take the form
\begin{align}
 \frac{r_{\rm ph}}{M}
 &=3+1.78834935\,q^2+\mathcal O(q^4),
 \label{eq:rph-final}\\
 \frac{b_{\rm ph}}{M}
 &=3\sqrt3+0.90589866\,q^2+\mathcal O(q^4),
 \label{eq:bph-final}\\
 \frac{r_{\rm MBO}}{M}
 &=4+1.10499638\,q^2+\mathcal O(q^4),
 \label{eq:rmbo-final}\\
 \frac{r_{\rm ISCO}}{M}
 &=6+0.38726014\,q^2+\mathcal O(q^4),
 \label{eq:risco-final}\\
 M\Omega_{\rm ISCO}
 &=\frac{1}{6\sqrt6}
 \left[1-0.137423046\,q^2+\mathcal O(q^4)\right],
 \label{eq:omega-isco-final}\\
 \frac{r_{3:2}}{M}
 &=10.8-1.79115679\,q^2+\mathcal O(q^4).
 \label{eq:r32-final}
\end{align}
The leading shifts are not uniform: the photon sphere, critical impact parameter, MBO, and ISCO move outward, whereas the ISCO frequency and the geometric $3{:}2$ radius decrease.

\begin{table}[t]
\centering
\caption{Leading ADM-normalized perturbative coefficients. The second column is the Schwarzschild value $X_0$. The third column is $C_q$ in $X=X_0+C_q q^2+\mathcal O(q^4)$; for the ISCO frequency the additive coefficient corresponding to \cref{eq:omega-isco-final} is shown. The last column rewrites the same result as $X=X_0+C_\alpha(\alpha-\alpha_c)+\cdots$.}
\label{tab:coefficients}
\begin{tabular}{lccc}
\toprule
Observable & $X_0$ & $C_q$ & $C_\alpha$\\
\midrule
$r_{\mathrm{ph}}/M$ & $3$ & $1.78834935$ & $0.16168540$\\
$b_{\mathrm{ph}}/M$ & $3\sqrt3$ & $0.90589866$ & $0.08190267$\\
$r_{\mathrm{MBO}}/M$ & $4$ & $1.10499638$ & $0.09990318$\\
$r_{\mathrm{ISCO}}/M$ & $6$ & $0.38726014$ & $0.03501235$\\
$M\Omega_{\mathrm{ISCO}}$ & $0.068041382$ & $-0.009350454$ & $-0.00084538$\\
$r_{3:2}/M$ & $10.8$ & $-1.79115679$ & $-0.16193922$\\
\bottomrule
\end{tabular}
\end{table}

\subsection{Derived timing and optical diagnostics}\label{subsec:derived-diagnostics}

The same metric correction also fixes the ISCO energetics and the optical diagnostics. At the shifted ISCO we obtain
\begin{align}
 E_{\rm ISCO}
 &=0.942809042-0.00217311\,q^2+\mathcal O(q^4),
 \label{eq:Eisco-final}\\
 \frac{L_{\rm ISCO}}{M}
 &=3.464101615-0.26537152\,q^2+\mathcal O(q^4),
 \label{eq:Lisco-final}\\
 \eta
 &=0.057190958+0.00217311\,q^2+\mathcal O(q^4).
 \label{eq:eta-final}
\end{align}
At the shifted $3{:}2$ frequency-ratio radius the two epicyclic frequencies become
\begin{align}
 M\Omega_\phi(r_{3:2})
 &=0.02817503+0.00579981\,q^2+\mathcal O(q^4),
 \label{eq:ophi-r32-final}\\
 M\Omega_r(r_{3:2})
 &=0.01878335+0.00386654\,q^2+\mathcal O(q^4),
 \label{eq:or-r32-final}
\end{align}
Their ratio is still $3/2$ at this order, but both frequencies are shifted upward.

For the photon orbit, \cref{eq:omega-photon,eq:lyapunov-photon} give
\begin{align}
 M\Omega_{\rm ph}
 &=\frac{1}{3\sqrt3}\left[1-0.17434028\,q^2+\mathcal O(q^4)\right],
 \label{eq:omega-ph-final}\\
 M\lambda_{\rm L}
 &=\frac{1}{3\sqrt3}\left[1+0.84223468\,q^2+\mathcal O(q^4)\right].
 \label{eq:lambda-final}
\end{align}
The circular photon orbit therefore moves outward and has a slightly lower angular frequency, while its radial instability grows. The corresponding geometric shadow diameter and area are
\begin{align}
 \frac{d_{\rm sh}}{M}
 &=6\sqrt3+1.81179732\,q^2+\mathcal O(q^4),
 \label{eq:dshadow-final}\\
 \frac{A_{\rm sh}}{M^2}
 &=27\pi\left[1+0.34868056\,q^2+\mathcal O(q^4)\right].
 \label{eq:Ashadow-final}
\end{align}
Finally, the strong-deflection logarithmic slope is
\begin{equation}
 \bar a=1-1.01657496\,q^2+\mathcal O(q^4).
 \label{eq:abar-final}
\end{equation}
The comparatively large fractional change in $\bar a$ reflects the opposite shifts of $\Omega_{\rm ph}$ and $\lambda_{\rm L}$.

\begin{table}[t]
\centering
\caption{Selected derived observables written as $X=X_0(1+S_q q^2)+\mathcal O(q^4)$. For $\eta$ the Schwarzschild reference is $1-2\sqrt2/3$.}
\label{tab:derived-coefficients}
\begin{tabular}{lc}
\toprule
Observable & $S_q$\\
\midrule
$E_{\rm ISCO}$ & $-0.002304932$\\
$L_{\rm ISCO}/M$ & $-0.07660616$\\
$\eta$ & $+0.03799745$\\
$\Omega_\phi(r_{3:2})$ & $+0.2058492$\\
$\Omega_r(r_{3:2})$ & $+0.2058490$\\
$\Omega_{\rm ph}$ & $-0.17434028$\\
$\lambda_{\rm L}$ & $+0.84223468$\\
$d_{\rm sh}$ & $+0.17434028$\\
$A_{\rm sh}$ & $+0.34868056$\\
$\bar a$ & $-1.01657496$\\
\bottomrule
\end{tabular}
\end{table}

For the plots we use the fractional shift relative to Schwarzschild,
\begin{equation}
 \delta_X(q)\equiv
 \frac{X(q)-X(0)}{X(0)}\times100\%.
 \label{eq:fractional-shift}
\end{equation}
Positive and negative values correspond, respectively, to an increase or decrease relative to Schwarzschild. All neutral-geodesic shifts shown here start at $\Ord(q^2)$.

\begin{figure*}[ht!]
\centering
\includegraphics[width=0.9\linewidth]{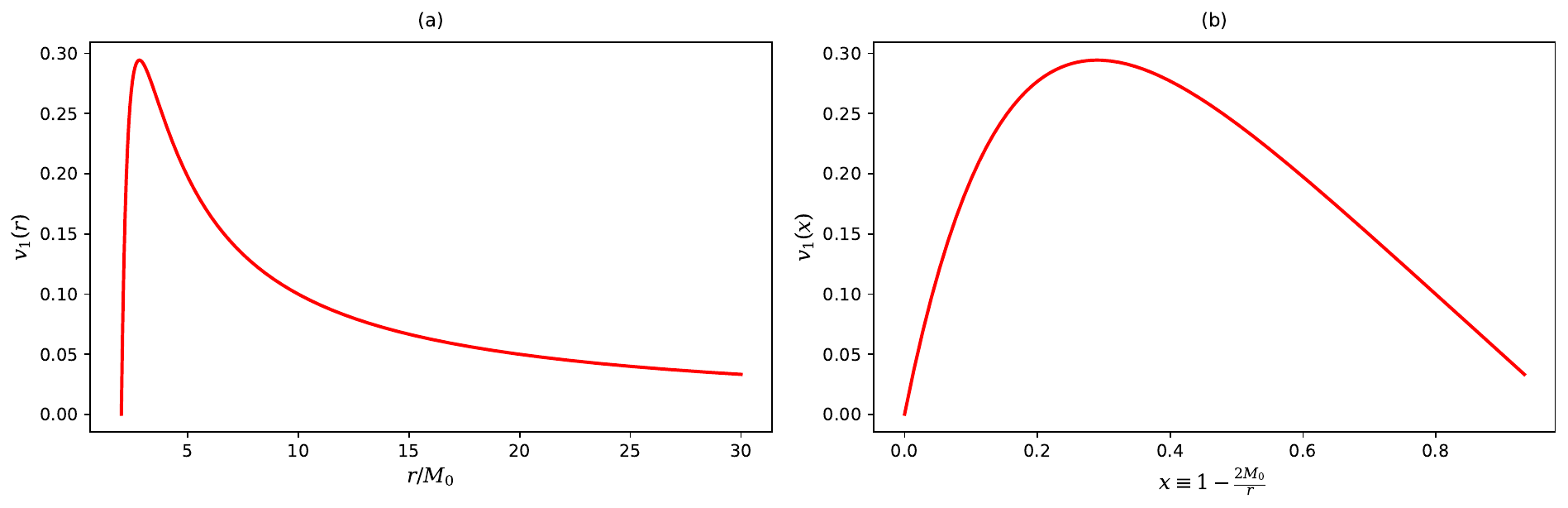}
\caption{Fundamental nodeless vector zero mode at the Schwarzschild bifurcation. The left panel uses the areal radius and the right panel the compact coordinate. The normalization is $rv_1/M_0\to1$ at infinity.}
\label{fig:zero-mode}
\end{figure*}

For reference, $q=0.1$ gives $\lambda/M^2\simeq4.79235$ at leading order, corresponding to $(\alpha-\alpha_c)/\alpha_c\simeq2.36\%$. This is a measure of distance from the bifurcation, not an estimate of the truncation error. At the same amplitude, $r_{\rm ph}$ increases by about $0.60\%$, $b_{\rm ph}$ (and therefore the shadow diameter) by $0.17\%$, $r_{\rm MBO}$ by $0.28\%$, and $r_{\rm ISCO}$ by $0.065\%$, while $r_{3:2}$ decreases by about $0.17\%$. These numbers only set the scale of the leading correction; the range over which the quadratic approximation remains accurate must be established by comparison with the nonlinear branch.

The finite-amplitude figures use the illustrative range $0\le q\le0.10$. It is chosen only to display the local quadratic trends and should not be read as a validity bound for the full branch. Summary plots are shown against $q^2$ where this makes the perturbative scaling explicit.

\begin{figure*}[ht!]
\centering
\includegraphics[width=0.9\linewidth]{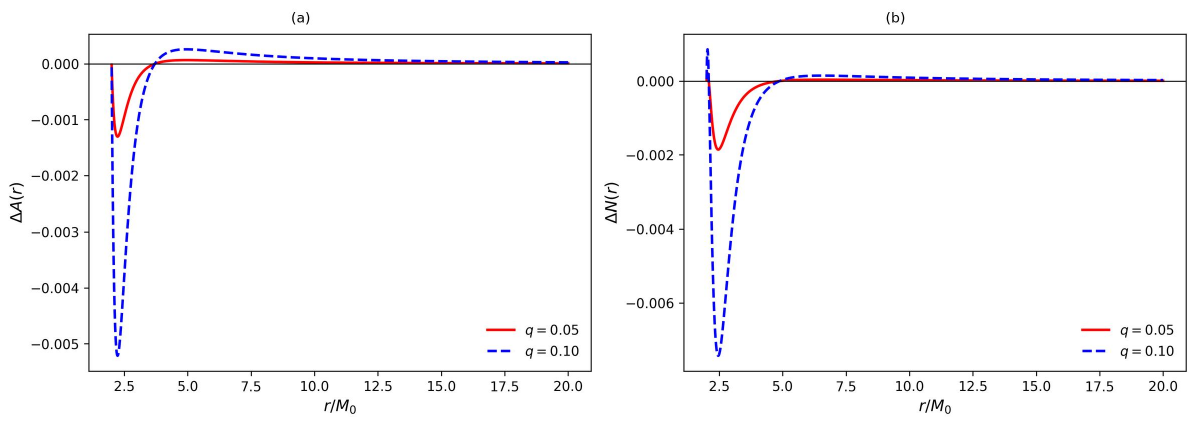}
\caption{Second-order metric deviations: (a) $\Delta A=q^2a(r)$ and (b) $\Delta N=q^2n(r)$. Red solid and blue dashed curves correspond to $q=0.05$ and $q=0.10$, respectively. The deviations are concentrated in the strong-field region and vanish asymptotically.}
\label{fig:metric-corrections}
\end{figure*}

\begin{figure*}[ht!]
\centering
\includegraphics[width=0.95\linewidth]{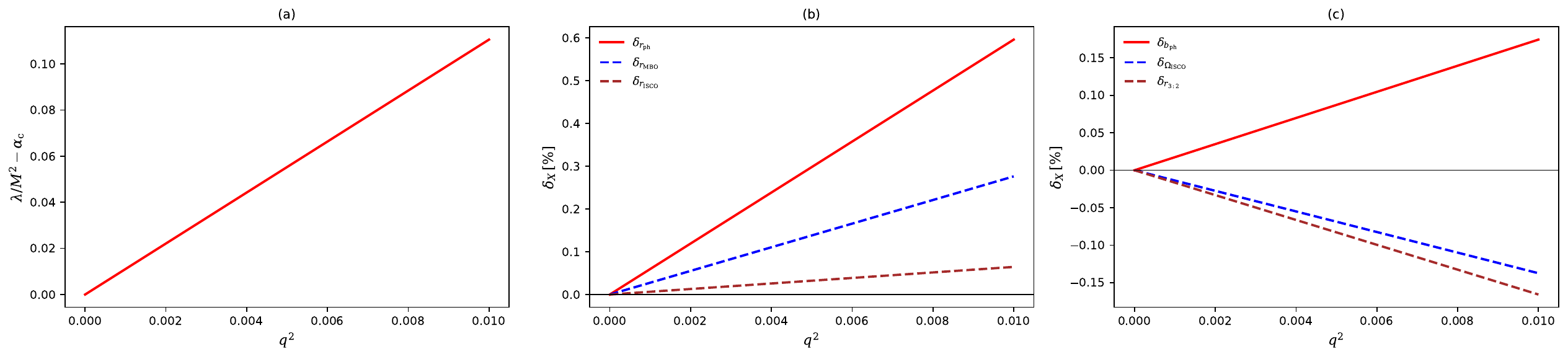}
\caption{Near-bifurcation scaling with $q^2$. (a) Shift of $\alpha=\lambda/M^2$ from the bifurcation value. (b) Fractional changes of the main orbital radii. (c) Fractional changes of the critical impact parameter and selected frequency/radius scales.}
\label{fig:branch-orbital-summary}
\end{figure*}

\begin{figure*}[ht!]
\centering
\includegraphics[width=0.95\linewidth]{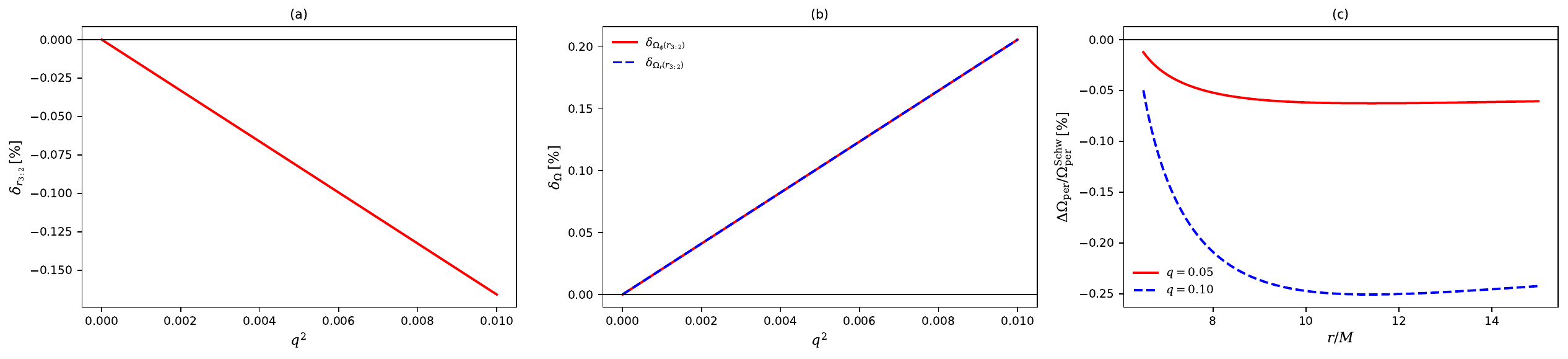}
\caption{Timing diagnostics. (a) Fractional shift of the geometric $3{:}2$ frequency-ratio radius. (b) Fractional shifts of $\Omega_\phi$ and $\Omega_r$ evaluated at the shifted radius. (c) Relative change of the periastron-precession frequency for $q=0.05$ (red solid) and $q=0.10$ (blue dashed). No QPO model is assumed.}
\label{fig:timing-summary}
\end{figure*}

\begin{figure*}[ht!]
\centering
\includegraphics[width=0.9\linewidth]{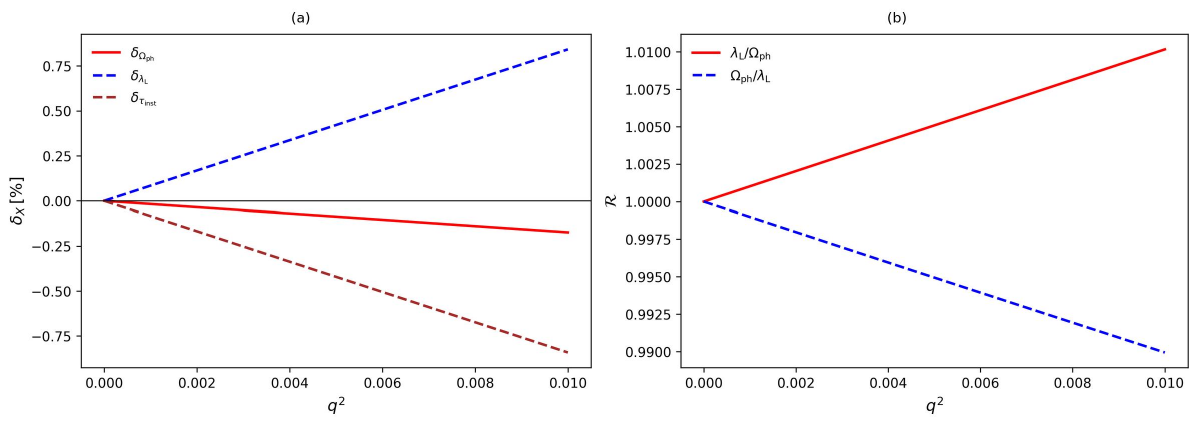}
\caption{Photon-orbit instability. (a) Fractional changes of the photon-orbit angular frequency, Lyapunov exponent, and local instability time. (b) Ratios $\lambda_{\rm L}/\Omega_{\rm ph}$ and $\Omega_{\rm ph}/\lambda_{\rm L}$, normalized to their Schwarzschild values.}
\label{fig:photon-instability-summary}
\end{figure*}

\begin{figure*}[ht!]
\centering
\includegraphics[width=0.9\linewidth]{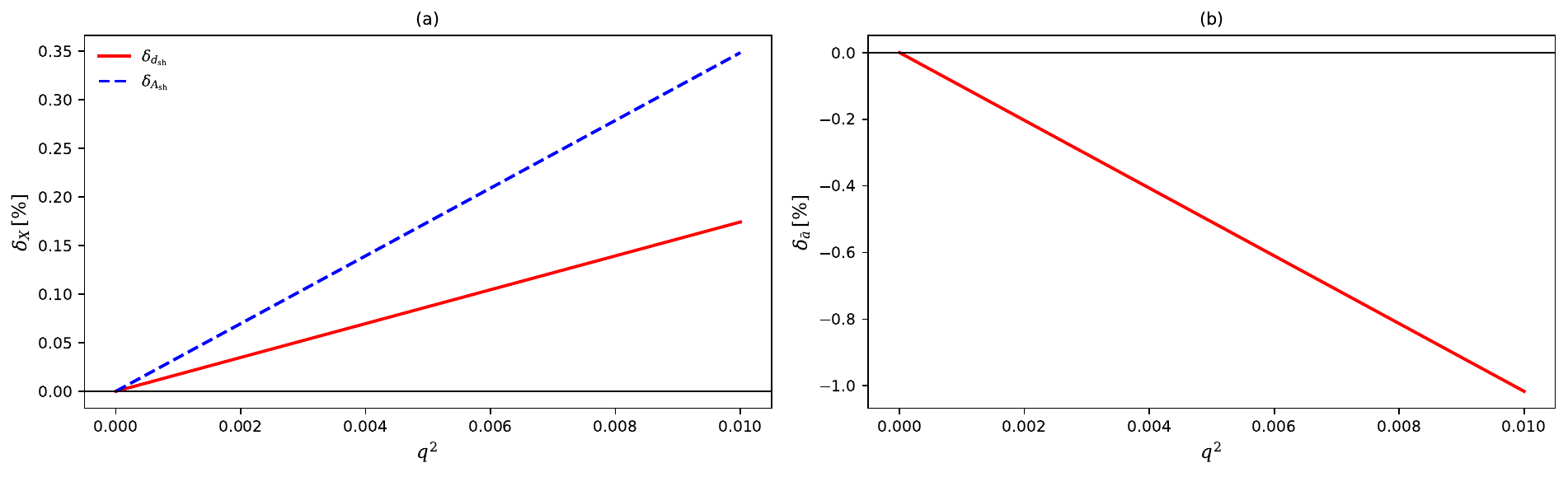}
\caption{Optical diagnostics. (a) Fractional changes of the geometric shadow diameter and area. (b) Fractional change of the logarithmic strong-deflection coefficient $\bar a$. The regular coefficient $\bar b$ is not included.}
\label{fig:shadow-lensing-compact}
\end{figure*}

\begin{figure}[t]
\centering
\includegraphics[width=0.98\linewidth]{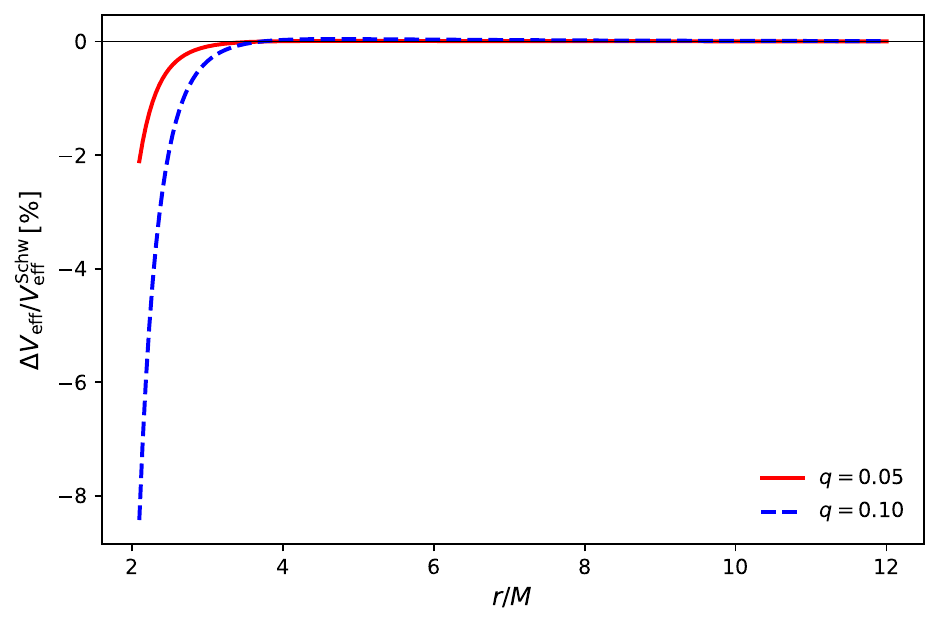}
\caption{Relative correction to the timelike effective potential for $q=0.05$ (red solid) and $q=0.10$ (blue dashed), shown after subtracting the Schwarzschild value.}
\label{fig:veff-profile}
\end{figure}

\begin{figure*}[ht!]
\centering
\includegraphics[width=0.9\linewidth]{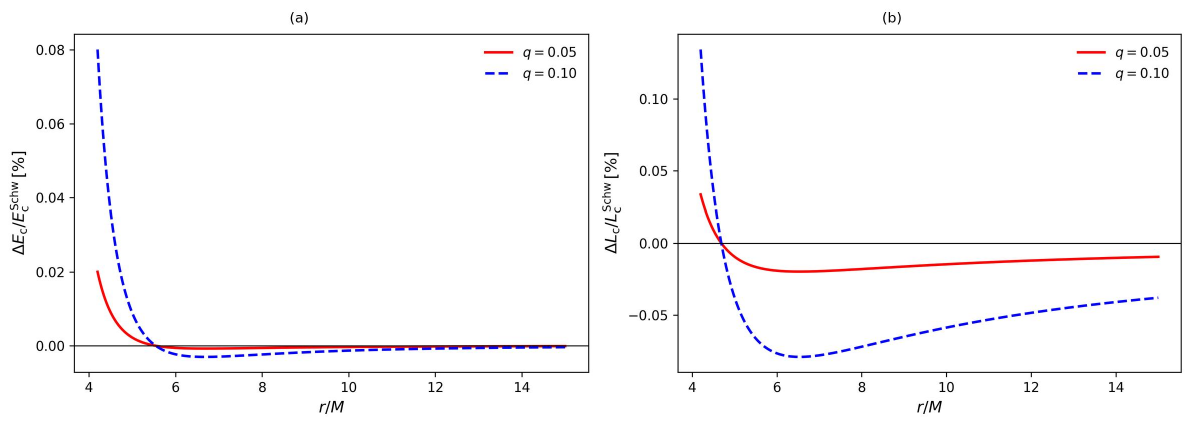}
\caption{Relative corrections to (a) the circular-orbit energy and (b) angular momentum for $q=0.05$ (red solid) and $q=0.10$ (blue dashed).}
\label{fig:EL-profile}
\end{figure*}

\begin{figure*}[ht!]
\centering
\includegraphics[width=0.9\linewidth]{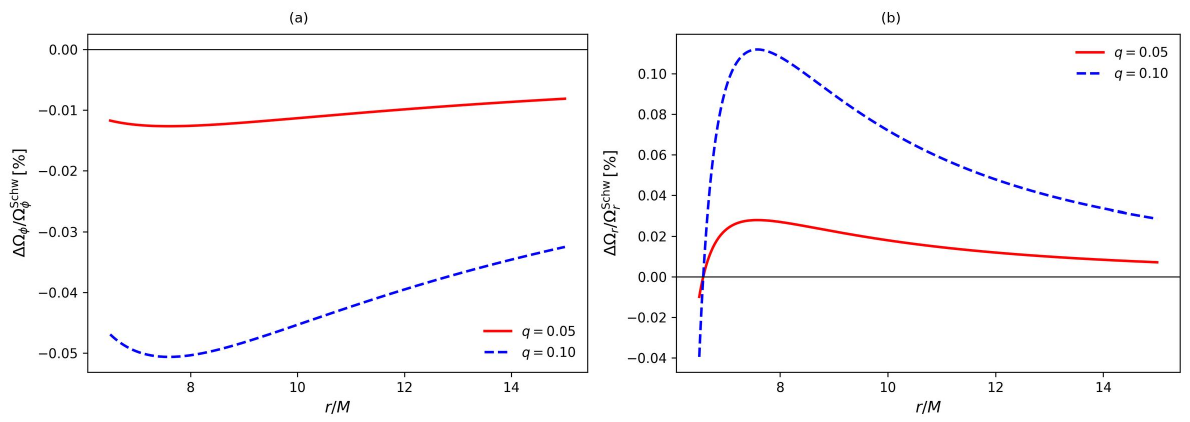}
\caption{Relative corrections to (a) the orbital frequency and (b) the radial epicyclic frequency for $q=0.05$ (red solid) and $q=0.10$ (blue dashed).}
\label{fig:freq-profile}
\end{figure*}

\begin{figure}[t]
\centering
\includegraphics[width=0.95\linewidth]{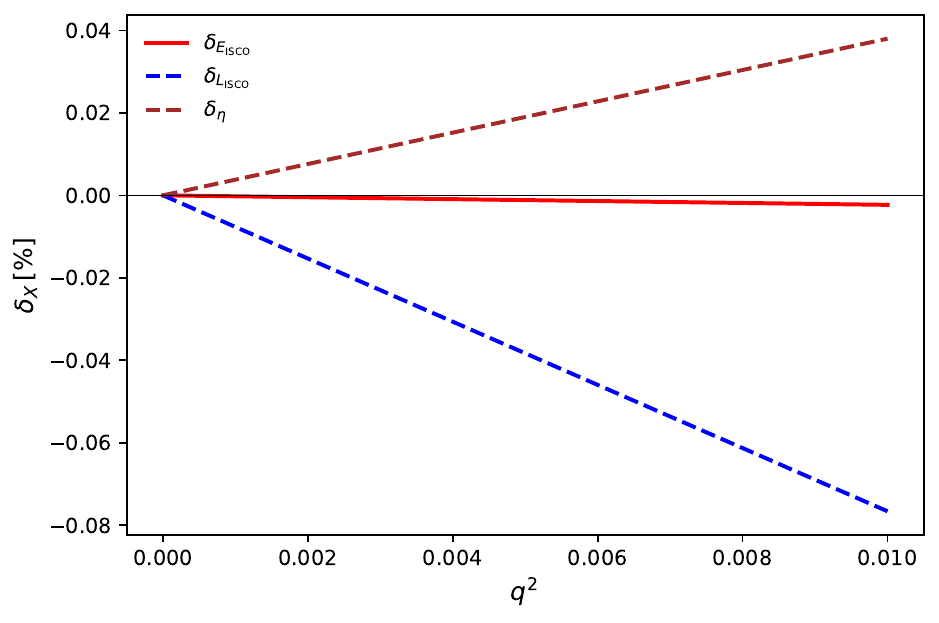}
\caption{Fractional changes of the ISCO energy, angular momentum, and geodesic binding-efficiency proxy $\eta=1-E_{\rm ISCO}$.}
\label{fig:isco-energetics}
\end{figure}

Figures~\ref{fig:zero-mode}--\ref{fig:shadow-lensing-compact} follow the calculation from the zero mode to metric backreaction and then to orbital, timing, and optical quantities. Figures~\ref{fig:veff-profile}--\ref{fig:isco-energetics} show the corresponding radial profiles and ISCO energetics.

Figure~\ref{fig:zero-mode} shows the mode that triggers the electric branch. It has no radial nodes, identifying it as the fundamental rather than an excited solution. The field vanishes at the horizon, is concentrated in the inner region, and approaches the $1/r$ tail fixed by Eq.~\eqref{eq:v-normalization}. This localization is expected because the curvature source is strongest near the black hole. Since $v_1$ is also the source of the second-order metric equations, the shape of the zero mode already indicates where the leading geometric corrections will be largest.

The metric response is shown in Fig.~\ref{fig:metric-corrections}. Both corrections are largest close to the horizon and decay quickly with radius, in accord with the absence of an $\Ord(q^2)$ ADM-mass shift. Doubling $q$ from $0.05$ to $0.10$ increases the curves by a factor of about four, as required by the perturbative scaling. The sign change away from the innermost region also shows that the deformation cannot be mimicked by a simple change of the Schwarzschild mass: the redshift and radial metric functions are modified differently across the strong-field region.

Figure~\ref{fig:branch-orbital-summary} makes the relative size of the leading effects easier to compare. The branch parameter increases linearly with $q^2$, as fixed by the positive Fredholm coefficient. Among the orbital radii, the photon sphere has the largest fractional displacement, the MBO shifts less, and the ISCO radius is only weakly affected. The critical impact parameter increases, whereas the ISCO frequency and the geometric $3{:}2$ radius decrease. The different signs and magnitudes reflect the fact that null and timelike circular-orbit conditions involve different combinations of $A$ and its derivatives; the deformation is therefore not a simple rescaling of all strong-field length scales.

The timing quantities in Fig.~\ref{fig:timing-summary} separate a shift of the preferred radius from a shift of the local frequencies. The geometric $3{:}2$ radius moves inward, while $\Omega_\phi$ and $\Omega_r$ evaluated at that new radius both increase. Their fractional changes are equal at the retained order because the defining ratio is fixed to $3/2$. The periastron-precession correction in panel (c) is instead evaluated as a radial profile and is negative over the interval shown. Together these trends show that the epicyclic sector is reshaped with radius rather than being multiplied by a single overall factor.

Figure~\ref{fig:photon-instability-summary} shows that the orbital and instability time scales change in opposite directions. The vectorized correction lowers $\Omega_{\rm ph}$ but raises $\lambda_{\rm L}$, so the photon orbit rotates slightly more slowly while radial departures from it grow faster. Accordingly, $\tau=1/\lambda_{\rm L}$ decreases and $\lambda_{\rm L}/\Omega_{\rm ph}$ increases. The shift of the photon-sphere radius alone would not reveal this change in local instability. These are geodesic quantities and should not be identified with the full coupled EvGB quasinormal spectrum.

The optical consequences are summarized in Fig.~\ref{fig:shadow-lensing-compact}. Since $b_{\rm ph}$ increases, the geometric shadow diameter and area both increase; at leading order the area has twice the fractional response of the diameter. The coefficient $\bar a$ changes in the opposite direction because it depends on the ratio $\Omega_{\rm ph}/\lambda_{\rm L}$. Its larger fractional shift concerns the local logarithmic part of the deflection law only. A complete lensing prediction would additionally require $\bar b$, a source--observer configuration, and any relevant propagation or emission model.

The remaining figures show the same corrections as radial profiles and at the ISCO, making clear where the $\mathcal O(q^2)$ response is concentrated.

Figure~\ref{fig:veff-profile} shows that the correction to the timelike effective potential is concentrated near the inner circular-orbit region and becomes negligible farther out. The $q=0.10$ curve is about four times the $q=0.05$ curve, again confirming the expected $q^2$ scaling. Although the absolute change in the potential is small, the MBO and ISCO depend on its derivatives, so a localized deformation can still move the corresponding radii. The rapid falloff with $r$ also explains the much weaker response of distant circular orbits.

The circular-orbit energetics in Fig.~\ref{fig:EL-profile} are most sensitive near the inner edge of the allowed domain, where the denominator $2A-rA'$ is small and the metric varies rapidly. Farther out, the energy correction crosses zero and quickly becomes negligible. The angular-momentum correction remains negative over a broader interval before decaying. Thus $E_c$ and $L_c$ do not respond in the same proportion; in the stable-orbit region the angular momentum is the more sensitive of the two, consistent with the ISCO coefficients below.

The frequency profiles in Fig.~\ref{fig:freq-profile} show a clear difference between azimuthal and radial dynamics. At fixed radius the orbital frequency is reduced across the interval shown. The correction to $\Omega_r$ changes sign in the inner stable-orbit region, becomes positive farther out, and then decays. This stronger radial structure is expected because $\Omega_r$ depends on second derivatives of the metric. Very close to the ISCO the fractional correction is not a useful diagnostic because the Schwarzschild radial frequency tends to zero. Away from that boundary, the profile displays the change in the radial restoring force that ultimately shifts the geometric $3{:}2$ radius.

At the ISCO, Fig.~\ref{fig:isco-energetics} shows a very small decrease in $E_{\rm ISCO}$ and a larger fractional decrease in $L_{\rm ISCO}$. Since $\eta=1-E_{\rm ISCO}$, the energy shift gives a small positive change in the geodesic binding-efficiency proxy. The straight lines in $q^2$ simply reflect the order at which the metric backreaction enters. Within the geodesic model the vectorized ISCO is therefore slightly more bound and requires less angular momentum than in Schwarzschild. The quantity $\eta$ should not be interpreted as a radiative efficiency without an accretion model.


\section{Discussion and limitations}\label{sec:discussion}

The calculation is local to the Schwarzschild bifurcation. Its small-amplitude assumptions are
\begin{equation}
 |\varepsilon|\ll1,
 \qquad
 |\alpha_{\mathrm{ADM}}-\ac|\ll1.
 \label{eq:validity}
\end{equation}
Accordingly, the series should not be extrapolated to large $q$ or far along the nonlinear electric branch. The sign of $\widehat\alpha_2$ is consistent with the known branch direction: the static spherical electric solutions bifurcate near $\lambda/M^2\simeq4.682$ and continue toward larger $\lambda/M^2$ \citep{BartonEtAl2021,KleihausKunz2026}. The curves up to $q=0.10$ illustrate the quadratic terms only. Establishing a quantitative validity interval requires a point-by-point comparison with the full numerical solutions.

The local calculation has several independent consistency checks. Shooting and collocation give the same fundamental eigenvalue, the reduced second-order equations satisfy independent components of the full covariant field equations, and the $\Ord(\varepsilon^2)$ ADM-mass correction vanishes through an exact total-derivative identity. These tests support the perturbative derivation but are not a substitute for comparison with the nonlinear branch.

We have also restricted the geometry to the static spherical sector. With rotation, prograde and retrograde orbits split, $\Omega_\theta$ no longer equals $\Omega_\phi$, nodal precession appears, and the shadow becomes noncircular. These effects could be treated with a joint expansion in spin and vector amplitude or by using the stationary numerical solutions of \citet{KleihausKunz2026}.

Dynamical viability is a separate issue. Ghost and gradient instabilities have been found in broad classes of tachyonic vectorization models \citep{SilvaCoatesRamazanoglu2022,PizzutiPombo2024}, and a recent no-go analysis extends this concern to a large set of stationary and axisymmetric GR backgrounds \citep{ChiangGarciaSaenzSang2026}. Those results are not, by themselves, a stability calculation for the particular nonlinear EvGB branch considered here; the assumptions of the general analyses must first be matched to this theory. We therefore make no claim about formation or dynamical stability. Addressing either question requires a dedicated perturbation or time-evolution study.

The optical quantities should be read with the same care. Equation~\eqref{eq:lyapunov-photon} describes the instability of a null geodesic; without a coupled gravitational--vector perturbation calculation it cannot be identified with the physical EvGB quasinormal-mode spectrum. Likewise, our lensing result contains only the local logarithmic coefficient $\bar a$. The regular coefficient $\bar b$, finite-distance effects, the source and observer geometry, plasma propagation, and radiative transfer are not included \citep{Bozza2002,CardosoEtAl2009,PerlickTsupko2022}. The geometric critical impact parameter therefore should not be confused with a complete prediction for a horizon-scale image, which also depends on the emitting plasma and observing setup \citep{MizunoEtAl2018,EHTM872019,EHTSgrA2022}.

\section{Conclusions}\label{sec:conclusions}

We have constructed the electric EvGB branch perturbatively in the immediate neighborhood of the Schwarzschild bifurcation and used it to calculate strong-field geodesic observables. The fundamental nodeless mode occurs at $\lambda/M^2=4.681746973049$. Independent shooting and collocation calculations give the same value to the precision quoted. At second order, the reduced metric equations agree with the unreduced covariant equations, and the mass equation can be written as an exact total derivative, giving a vanishing fixed-horizon ADM-mass correction. The third-order Fredholm condition fixes the local branch relation, $\lambda/M^2=\alpha_c+11.060673q^2+\cdots$.

The leading geodesic corrections have a clear pattern. The photon sphere, MBO, and ISCO move outward, but the geometric $3{:}2$ radius moves inward. The ISCO angular frequency decreases, whereas the orbital and radial epicyclic frequencies evaluated at the shifted $3{:}2$ radius increase. The circular photon orbit has a lower angular frequency and a larger Lyapunov exponent. The same competition gives $\bar a=1-1.016575q^2+\mathcal O(q^4)$ in the strong-deflection expansion. We also obtain the corresponding ISCO energy, angular momentum, binding-efficiency proxy, periastron-precession correction, and geometric shadow shifts.

These coefficients characterize the local geometry of the branch; they are not observational constraints and do not establish how the solutions form dynamically. Comparison with the full nonlinear metric is needed to determine how far the quadratic expansion can be trusted. Rotation and dynamical stability remain separate problems. In particular, a coupled gravitational--vector perturbation analysis is needed before relating the null-geodesic Lyapunov exponent to the physical quasinormal-mode spectrum.

\begin{acknowledgments}
S.M. and B.A. gratefully acknowledge support from Grant FL--10425067111 of the Agency of Innovative Developments of Uzbekistan.
\end{acknowledgments}

\appendix

\section{Reduced field equations and weakly nonlinear details}\label{app:weak-details}

This appendix records the intermediate formulas used in the weakly nonlinear calculation.

\subsection{Vector variation and reduced radial action}

Varying \cref{eq:action} with respect to $A_\nu$ gives
\begin{align}
 \delta_A S
 &=\frac{1}{16\pi}\int\dd^4x\sqrt{-g}
 \left[-2F^{\mu\nu}\delta F_{\mu\nu}
 +2\lambda A^\nu\GB\,\delta A_\nu\right]
 \nonumber\\
 &=\frac{1}{16\pi}\int\dd^4x\sqrt{-g}
 \left[4\nabla_\mu F^{\mu\nu}
 +2\lambda A^\nu\GB\right]\delta A_\nu,
 \label{eq:variation-vector}
\end{align}
after discarding a boundary term, which reproduces Eq.~\eqref{eq:vector-eom}. For the static spherical ansatz, the determinant and invariants are given in Eq.~\eqref{eq:basic-invariants}. Direct evaluation of the curvature scalars gives
\begin{align}
 R={}&-\frac{1}{r^2}\Big[
 2r^2N(\delta')^2-2r^2N\delta''
 -3r^2N'\delta'\nonumber\\
 &\quad+r^2N''-4rN\delta'+4rN'
 +2N-2\Big],
 \label{eq:ricci-reduced}\\[1ex]
 \GB={}&\frac{4}{r^2}\Big[
 2N^2(\delta')^2-2N^2\delta''
 -5NN'\delta'\nonumber\\
 &\quad+NN''-2N(\delta')^2+2N\delta''
 +(N')^2\nonumber\\
 &\quad+3N'\delta'-N''\Big].
 \label{eq:GB-reduced}
\end{align}
After integration over $t$ and the two-sphere, the radial Lagrangian density, up to an overall constant, is
\begin{equation}
 \mathcal L_{1\mathrm D}
 =\ee^{-\delta}r^2
 \left[
 R+2\ee^{2\delta}(V')^2
 -\lambda\frac{\ee^{2\delta}V^2}{N}\GB
 \right].
 \label{eq:reduced-lagrangian}
\end{equation}
Since \cref{eq:ricci-reduced,eq:GB-reduced} contain second radial derivatives, the radial equations follow from the generalized Euler--Lagrange operator
\begin{equation}
 \frac{\partial\mathcal L_{1\mathrm D}}{\partial q}
 -\frac{\dd}{\dd r}
 \frac{\partial\mathcal L_{1\mathrm D}}{\partial q'}
 +\frac{\dd^2}{\dd r^2}
 \frac{\partial\mathcal L_{1\mathrm D}}{\partial q''}=0,
 \qquad q\in\{N,\delta\}.
 \label{eq:generalized-EL}
\end{equation}
The higher radial derivatives cancel in the final field equations.

\subsection{Near-horizon zero-mode expansion}

Near the horizon, with $x=r-2M_0$, the regular solution has the expansion
\begin{equation}
 v_1=a_1x+a_2x^2+a_3x^3+\cdots,
 \label{eq:v-horizon-series}
\end{equation}
where \cref{eq:zero-mode-ode} gives
\begin{align}
 a_2&=-\frac{4M_0^2+3\lambda_{\mathrm c}}{8M_0^3}a_1,
 \label{eq:a2}\\
 a_3&=\frac{16M_0^4+32M_0^2\lambda_{\mathrm c}
 +3\lambda_{\mathrm c}^2}{64M_0^6}a_1.
 \label{eq:a3}
\end{align}
These coefficients are used to initialize the shooting integration just outside the horizon.

\subsection{Second-order backreaction equations}

At $\Ord(\varepsilon^2)$, the vector zero mode sources the metric. Expanding the generalized Euler--Lagrange equations obtained from \cref{eq:reduced-lagrangian} yields two first-order equations. For compactness, define
\begin{equation}
 \Delta(r)=r-2M_0,
 \qquad u(r)\equiv v_1(r).
 \label{eq:Delta-u}
\end{equation}
The mass-function equation is
\begin{align}
 m_2'={}&\frac{r^2}{2}(u')^2
 -\frac{8M_0\lambda_{\mathrm c}}{r}
 \left[(u')^2+uu''\right]
 \nonumber\\
 &+\frac{8M_0\lambda_{\mathrm c}(r+M_0)}{r^2\Delta}
 uu'
 +\frac{24M_0^2\lambda_{\mathrm c}(3M_0-2r)}
 {r^3\Delta^2}u^2,
 \label{eq:m2-prime}
\end{align}
and the redshift equation is
\begin{align}
 \delta_2'={}&\frac{8M_0\lambda_{\mathrm c}}
 {r^3\Delta^3}
 \Big\{
 r^2\Delta^2\left[(u')^2+uu''\right]
 -4M_0r\Delta\,uu'
 \nonumber\\
 &\hspace{4.3cm}
 +M_0(5r-6M_0)u^2
 \Big\}.
 \label{eq:delta2-prime}
\end{align}
The apparent powers of $\Delta^{-1}$ are removable for the regular zero mode. Numerically, one may eliminate $u''$ using
\begin{equation}
 u''=-\frac{2}{r}u'
 -\frac{24\lambda_{\mathrm c}M_0^2}{r^5\Delta}u,
 \label{eq:u-double-prime}
\end{equation}
combined with the horizon series \cref{eq:v-horizon-series}.

The fixed-horizon boundary conditions are given in Eq.~\eqref{eq:second-order-bcs}. The corresponding derivatives remain finite:
\begin{align}
 m_2'(2M_0)&=a_1^2\left(2M_0^2-\lambda_{\mathrm c}\right),
 \label{eq:m2-horizon}\\
 \delta_2'(2M_0)&=
 \frac{a_1^2\lambda_{\mathrm c}}{M_0^3}
 \left(M_0^2-3\lambda_{\mathrm c}\right).
 \label{eq:delta2-horizon}
\end{align}

The asymptotic expansions are given in Eqs.~\eqref{eq:m2-asymptotic} and \eqref{eq:delta2-asymptotic}.

\subsection{Analytic vanishing of the second-order mass correction}\label{app:mu2-identity}

The small numerical value of $\mu_2$ reflects an exact identity. Eliminating $u''$ from Eq.~\eqref{eq:m2-prime} with Eq.~\eqref{eq:u-double-prime} rewrites the mass source as
\begin{align}
 m_2'(r)
 =\frac{\dd}{\dd r}\Bigg[
 &\left(\frac{r^3-16\lambda_{\mathrm c}M_0}{2r}\right)uu'
 \nonumber\\
 &+\frac{12\lambda_{\mathrm c}M_0^2}
 {r^2(r-2M_0)}u^2
 \Bigg].
 \label{eq:m2-total-derivative}
\end{align}
For the regular horizon expansion $u=a_1(r-2M_0)+\mathcal O[(r-2M_0)^2]$, the quantity in square brackets is $\mathcal O(r-2M_0)$ and therefore vanishes at the horizon. At infinity, the normalization $u=M_0/r+\mathcal O(r^{-5})$ gives
\begin{equation}
 \left[\cdots\right]
 =-\frac{M_0^2}{2r}+\mathcal O(r^{-4})
 \longrightarrow0.
 \label{eq:m2-boundary-infinity}
\end{equation}
Hence
\begin{equation}
 m_2(\infty)-m_2(2M_0)=0.
 \label{eq:m2-integrated-identity}
\end{equation}
With $m_2(2M_0)=0$, Eq.~\eqref{eq:m2-integrated-identity} proves Eq.~\eqref{eq:mu2-exact}. No statement is implied about higher-order mass corrections on the nonlinear branch.

\subsection{Third-order Fredholm solvability}

The relation between the amplitude and the displacement from the bifurcation enters at $\Ord(\varepsilon^3)$. Expand the Gauss--Bonnet invariant as
\begin{equation}
 \GB=\GB_0+\varepsilon^2\GB_2+\Ord(\varepsilon^4),
 \label{eq:G-expansion}
\end{equation}
where direct expansion of \cref{eq:GB-reduced} gives
\begin{align}
 \GB_2={}&\frac{16M_0}{r^6}
 \Big[
 -2M_0r^2\delta_2''+5M_0r\delta_2'
 +r^3\delta_2''-r^2\delta_2'
 \nonumber\\
 &\hspace{3.3cm}
 +r^2m_2''-4rm_2'+6m_2
 \Big].
 \label{eq:G2}
\end{align}
Substitution into the exact vector equation gives
\begin{align}
 \mathcal L_{\mathrm c}[v_3]
 ={}&-\delta_2'u'
 -\Bigg[
 \frac{M_0^2\alpha_2^{(\mathrm H)}\GB_0}{2f}
 +\frac{\lambda_{\mathrm c}\GB_2}{2f}
 +\frac{\lambda_{\mathrm c}\GB_0m_2}{rf^2}
 \Bigg]u.
 \label{eq:v3-equation}
\end{align}
The operator \cref{eq:self-adjoint-operator} is self-adjoint with weight $r^2$. With horizon-fixed homogeneous boundary conditions for $v_3$ and with the $1/r$ part of $v_3$ removed by the definition of $\varepsilon$, the Fredholm condition is
\begin{align}
 0={}&\int_{2M_0}^{\infty}\dd r\,r^2u
 \Bigg\{\delta_2'u'
 +\Bigg[
 \frac{M_0^2\alpha_2^{(\mathrm H)}\GB_0}{2f}
 \nonumber\\[-1mm]
 &\qquad+\frac{\lambda_{\mathrm c}\GB_2}{2f}
 +\frac{\lambda_{\mathrm c}\GB_0m_2}{rf^2}
 \Bigg]u\Bigg\}.
 \label{eq:fredholm}
\end{align}
Therefore,
\begin{equation}
 \alpha_2^{(\mathrm H)}
 =-\frac{2\displaystyle\int_{2M_0}^{\infty}\dd r\,r^2
 \left[
 \delta_2'uu'
 +\frac{\lambda_{\mathrm c}\GB_2}{2f}u^2
 +\frac{\lambda_{\mathrm c}\GB_0m_2}{rf^2}u^2
 \right]}
 {M_0^2\displaystyle\int_{2M_0}^{\infty}\dd r\,r^2
 \frac{\GB_0}{f}u^2}.
 \label{eq:alpha2-solvability}
\end{equation}
The sign of this coefficient fixes the local direction of the branch.

\subsection{Independent check against the unreduced metric equations}\label{subsec:full-equation-check}

We also checked the symmetry-reduced equations by expanding the full covariant EvGB metric equations of Ref.~\citep{BartonEtAl2021} directly about Schwarzschild. For the quadratic coupling used here, define the second-order covariant residual
\begin{equation}
 \mathcal E_{\mu\nu}^{(2)}
 \equiv
 \left[
 G_{\mu\nu}-\frac12 T^{(A)}_{\mu\nu}
 +T^{(\mathrm{GB})}_{\mu\nu}
 \right]_{\Ord(\varepsilon^2)} .
 \label{eq:full-covariant-residual}
\end{equation}
The tensor $T^{(\mathrm{GB})}_{\mu\nu}$ was evaluated in the double-dual-Riemann form of the unreduced equations, including the contribution from the explicit metric dependence of $A_\rho A^\rho$. This calculation was carried out independently of the one-dimensional action. Substituting
\begin{equation}
 n=-\frac{2m_2}{r},
 \qquad
 a=-\frac{2m_2}{r}-2f\delta_2,
 \label{eq:full-check-an}
\end{equation}
together with the zero-mode equation and the two second-order radial source equations, symbolic simplification gives
\begin{equation}
 \mathcal E_{tt}^{(2)}=
 \mathcal E_{rr}^{(2)}=
 \mathcal E_{\theta\theta}^{(2)}=0
 \label{eq:full-equation-check}
\end{equation}
identically, with $\mathcal E_{\phi\phi}^{(2)}=\sin^2\theta\,\mathcal E_{\theta\theta}^{(2)}$ by spherical symmetry. Thus the reduced radial equations reproduce the independent covariant metric components at the order used here. The symbolic check is included with the reproducibility material.

\subsection{Algebraic checks}

For $\delta=0$ and $N=f=1-2M_0/r$, Eqs.~\eqref{eq:ricci-reduced} and \eqref{eq:GB-reduced} give
\begin{equation}
 R=0,
 \qquad
 \GB=\frac{48M_0^2}{r^6},
\end{equation}
which verifies Eq.~\eqref{eq:GB-schwarzschild}. In the limit $\lambda\rightarrow0$, Eqs.~\eqref{eq:m2-prime} and \eqref{eq:delta2-prime} reduce to
\begin{equation}
 m_2'=\frac{r^2}{2}(u')^2,
 \qquad
 \delta_2'=0.
 \label{eq:maxwell-check}
\end{equation}
As a formal normalization check, consider the decoupled $\lambda=0$ Einstein--Maxwell theory and set $u=M_0/r$. Then $m_2'=M_0^2/(2r^2)$, giving the expected Reissner--Nordstr\"om-type $1/r^2$ correction for the normalization in Eq.~\eqref{eq:action}. This Coulomb profile is not the EvGB zero mode and does not obey the horizon condition $u(2M_0)=0$ used in the coupled problem; $U(1)$ gauge symmetry is restored at $\lambda=0$. The comparison is only an algebraic check of normalization.

\section{Perturbative derivation of orbital and timing shifts}\label{app:orbital-derivations}

For completeness, this appendix gives the second-order shift formulas used for the numerical coefficients in Sec.~\ref{sec:results}.

\subsection{Photon-sphere and shadow shifts}\label{subsec:photon-shifts}

Using the generic root-displacement formula, Eq.~\eqref{eq:root-shift}, the photon-sphere correction is
\begin{equation}
 \delta r_{\mathrm{ph}}
 =\frac{3M_0}{2}
 \left[ra'-2a\right]_{r=3M_0}.
 \label{eq:delta-rph}
\end{equation}
Thus
\begin{equation}
 r_{\mathrm{ph}}=3M_0
 +\varepsilon^2\delta r_{\mathrm{ph}}
 +\Ord(\varepsilon^4).
 \label{eq:rph-expansion}
\end{equation}
Because $r/\sqrt{A}$ is stationary at the photon sphere, the shift of the critical impact parameter receives no explicit contribution from $\delta r_{\mathrm{ph}}$ at this order. One finds
\begin{equation}
 \delta b_{\mathrm{ph}}
 =-\frac{9\sqrt{3}}{2}M_0\,a(3M_0),
 \label{eq:delta-bph}
\end{equation}
with
\begin{equation}
 b_{\mathrm{ph}}=3\sqrt{3}M_0
 +\varepsilon^2\delta b_{\mathrm{ph}}
 +\Ord(\varepsilon^4).
 \label{eq:bph-expansion}
\end{equation}

\subsection{Marginally bound-orbit shift}\label{subsec:mbo-shift}

Expanding \cref{eq:mbo-condition} around the Schwarzschild root $r=4M_0$ and applying \cref{eq:root-shift} gives
\begin{equation}
 \delta r_{\rm MBO}
 =32M_0^2 a'(4M_0).
 \label{eq:delta-rmbo}
\end{equation}
Thus
\begin{equation}
 r_{\rm MBO}=4M_0+\varepsilon^2\delta r_{\rm MBO}+\Ord(\varepsilon^4).
 \label{eq:rmbo-expansion}
\end{equation}

\subsection{ISCO radius and frequency}\label{subsec:isco-shifts}

Expanding \cref{eq:isco-condition} gives
\begin{equation}
 \mathcal I=\mathcal I_0+\varepsilon^2\mathcal I_2+\Ord(\varepsilon^4),
 \label{eq:I-expansion}
\end{equation}
where
\begin{equation}
 \mathcal I_2
 =3(fa'+af')-4rf'a'+r(fa''+af'').
 \label{eq:I2}
\end{equation}
Since $\mathcal I_0'(6M_0)=1/(108M_0^2)$, the ISCO shift is
\begin{equation}
 \delta r_{\mathrm{ISCO}}
 =-6M_0
 \left[a+12M_0a'+72M_0^2a''\right]_{r=6M_0}.
 \label{eq:delta-risco}
\end{equation}
The orbital frequency at the shifted ISCO satisfies
\begin{align}
 \Omega_{\mathrm{ISCO}}
 &={\frac{1}{6\sqrt{6}M_0}}
 \Big[1+\varepsilon^2\mathcal C_{\Omega}^{(\mathrm H)}
 +\Ord(\varepsilon^4)\Big],
 \label{eq:Omega-isco-expansion}\\
 \mathcal C_{\Omega}^{(\mathrm H)}
 &=9M_0a'(6M_0)
 -\frac{\delta r_{\mathrm{ISCO}}}{4M_0}.
 \label{eq:C-Omega}
\end{align}

\subsection{Epicyclic-frequency correction}\label{subsec:epicyclic-shifts}

Define
\begin{equation}
 \mathcal K_0
 =f''-\frac{2(f')^2}{f}+\frac{3f'}{r}.
 \label{eq:K0}
\end{equation}
Using $N=f+\varepsilon^2n$ and \cref{eq:Omega-r}, we obtain
\begin{align}
 \Omega_r^2
 &=\Omega_{r,0}^2+\varepsilon^2\Xi_r
 +\Ord(\varepsilon^4),
 \label{eq:Omega-r-expansion}\\
 \Xi_r
 &=\frac12\Bigg\{
 n\mathcal K_0
 +f\Bigg[
 a''-\frac{4f'a'}{f}\nonumber\\[-1mm]
 &\qquad+\frac{2(f')^2a}{f^2}
 +\frac{3a'}{r}
 \Bigg]\Bigg\}.
 \label{eq:Xi-r}
\end{align}
Similarly,
\begin{equation}
 \Omega_\phi^2
 =\frac{M_0}{r^3}
 +\varepsilon^2\frac{a'}{2r}
 +\Ord(\varepsilon^4).
 \label{eq:Omega-phi-expansion}
\end{equation}
Away from the ISCO, where $\Omega_{r,0}\neq0$,
\begin{equation}
 \delta\Omega_r=\frac{\Xi_r}{2\Omega_{r,0}}.
 \label{eq:delta-Omega-r}
\end{equation}

\subsection{Shift of the geometric $3{:}2$ radius}

At second order,
\begin{equation}
 \mathcal H_2=\frac{2a'}{r}-9\Xi_r.
 \label{eq:H2}
\end{equation}
Since $\mathcal H_0'(r_{3:2}^{(0)})=-5M_0/[r_{3:2}^{(0)}]^4$, the frequency-ratio-radius shift is
\begin{equation}
 \delta r_{3:2}
 =\frac{[r_{3:2}^{(0)}]^4}{5M_0}
 \left[\frac{2a'}{r}-9\Xi_r\right]_{r=r_{3:2}^{(0)}}.
 \label{eq:delta-r32}
\end{equation}
An observational QPO analysis is left for future work, after the perturbative coefficients have been compared with the full numerical branch.

\subsection{ADM-normalized observables}\label{subsec:ADM-observables}

Let an orbital radius have the horizon-fixed expansion
\begin{equation}
 r_X=r_X^{(0)}+\varepsilon^2\delta r_X+\Ord(\varepsilon^4),
 \qquad
 x_X^{(0)}\equiv\frac{r_X^{(0)}}{M_0}.
 \label{eq:generic-radius}
\end{equation}
Using \cref{eq:ADM-mass}, the physically normalized result is
\begin{equation}
 \frac{r_X}{\Minf}
 =x_X^{(0)}+\varepsilon^2
 \left[
 \frac{\delta r_X}{M_0}
 -x_X^{(0)}\frac{\mu_2}{M_0}
 \right]
 +\Ord(\varepsilon^4).
 \label{eq:radius-ADM-normalized}
\end{equation}
The same formula applies to $b_{\mathrm{ph}}$. For a frequency,
\begin{equation}
 \Minf\Omega_X
 =M_0\Omega_X^{(0)}
 +\varepsilon^2
 \left[M_0\delta\Omega_X+\mu_2\Omega_X^{(0)}\right]
 +\Ord(\varepsilon^4).
 \label{eq:frequency-ADM-normalized}
\end{equation}
Finally, \cref{eq:epsilon-alpha} converts the amplitude expansion into an expansion in the physical distance from the bifurcation.

\section{Derivation of the radial epicyclic frequency}\label{app:epicyclic-derivation}

Define
\begin{equation}
 \mathcal R(r)=\frac{E^2}{A}-1-\frac{L^2}{r^2},
 \qquad
 \dot r^2=\frac{\mathcal R}{B}.
 \end{equation}
At a circular orbit, $\mathcal R=\mathcal R'=0$. A perturbation $r=r_0+\delta r$ obeys
\begin{equation}
 \frac{\dd^2\delta r}{\dd\tau^2}
 +\omega_r^2\delta r=0,
 \qquad
 \omega_r^2=-\frac{\mathcal R''(r_0)}{2B(r_0)}.
 \end{equation}
Since $\dd t/\dd\tau=E/A$, the coordinate-time frequency is
\begin{equation}
 \Omega_r^2=\omega_r^2\frac{A^2}{E^2}.
 \end{equation}
Substituting the circular-orbit expressions \cref{eq:EL-circular} and simplifying yields \cref{eq:Omega-r}.

\bibliographystyle{apsrev4-2}
\bibliography{references}

@misc{MurodovEtAl2026EsGBQPO,
  author       = {Murodov, Sardor and Rahmatov, Bekzod and Umarov, Otabek
                  and Kurbaniyazov, Anvar and Javed, Faisal and Rayimbaev, Javlon},
  title        = {Circular-orbit dynamics and QPO constraints in static
                  Einstein--scalar--Gauss--Bonnet black holes},
  year         = {2026},
  eprint       = {2607.19805},
  archivePrefix= {arXiv},
  primaryClass = {gr-qc}
}

@article{AhmedMurodovEtAl2026NPB,
  author={Ahmed, Faizuddin and Murodov, Sardor and Rahmatov, Bekzod},
  title={Black holes in general relativity coupled with NEDs surrounded by PFDM: Thermodynamics, epicyclic oscillations, QPOs, and shadow},
  journal={Nuclear Physics B},
  volume={1029},
  pages={117556},
  year={2026},
  doi={10.1016/j.nuclphysb.2026.117556}
}

@article{MurodovEtAl2026KS,
  author = {{Murodov}, Sardor and {Egamberdiev}, Islom and {Umarov}, Otabek and {Rayimbaev}, Javlon and {Davletov}, Erkaboy and {Davletov}, Ikram},
        title = "{Probing quantum corrected Kazakov─Solodukhin black holes through QPOs and thin accretion disks}",
      journal = {European Physical Journal C},
         year = 2026,
        month = may,
       volume = {86},
       number = {5},
          eid = {490},
        pages = {490},
          doi = {10.1140/epjc/s10052-026-15636-w},
       adsurl = {https://ui.adsabs.harvard.edu/abs/2026EPJC...86..490M}
}

@article{MurodovRayimbaevAhmedov2023Universe,
  author={Murodov, Sardor and Rayimbaev, Javlon and Ahmedov, Bobomurat and Karimbaev, Eldor},
  title={Quasiperiodic Oscillations and Dynamics of Test Particles around Quasi- and Non-Schwarzschild Black Holes},
  journal={Universe},
  volume={9},
  number={9},
  pages={391},
  year={2023},
  doi={10.3390/universe9090391}
}

@article{KleihausKunz2026,
  author  = {Kleihaus, Burkhard and Kunz, Jutta},
  title   = {Stationary Einstein-vector-Gauss-Bonnet black holes},
  journal = {Physics Letters B},
  volume  = {879},
  pages   = {140640},
  year    = {2026},
  doi     = {10.1016/j.physletb.2026.140640},
  eprint  = {2604.04568},
  archivePrefix = {arXiv},
  primaryClass  = {gr-qc}
}

@article{BartonEtAl2021,
  author  = {Barton, Simon and Hartmann, Betti and Kleihaus, Burkhard and Kunz, Jutta},
  title   = {Spontaneously vectorized Einstein-Gauss-Bonnet black holes},
  journal = {Physics Letters B},
  volume  = {817},
  pages   = {136336},
  year    = {2021},
  doi     = {10.1016/j.physletb.2021.136336},
  eprint  = {2103.01651},
  archivePrefix = {arXiv},
  primaryClass  = {gr-qc}
}

@article{DonevaYazadjiev2018,
  author  = {Doneva, Daniela D. and Yazadjiev, Stoytcho S.},
  title   = {New Gauss-Bonnet black holes with curvature-induced scalarization in extended scalar-tensor theories},
  journal = {Physical Review Letters},
  volume  = {120},
  number  = {13},
  pages   = {131103},
  year    = {2018},
  doi     = {10.1103/PhysRevLett.120.131103}
}

@article{SilvaEtAl2018,
  author  = {Silva, Hector O. and Sakstein, Jeremy and Gualtieri, Leonardo and Sotiriou, Thomas P. and Berti, Emanuele},
  title   = {Spontaneous scalarization of black holes and compact stars from a Gauss-Bonnet coupling},
  journal = {Physical Review Letters},
  volume  = {120},
  number  = {13},
  pages   = {131104},
  year    = {2018},
  doi     = {10.1103/PhysRevLett.120.131104}
}

@article{MotohashiMukohyama2019,
  author  = {Motohashi, Hayato and Mukohyama, Shinji},
  title   = {Shape dependence of spontaneous scalarization},
  journal = {Physical Review D},
  volume  = {99},
  number  = {4},
  pages   = {044030},
  year    = {2019},
  doi     = {10.1103/PhysRevD.99.044030}
}

@article{CunhaEtAl2019,
  author  = {Cunha, Pedro V. P. and Herdeiro, Carlos A. R. and Radu, Eugen},
  title   = {Spontaneously scalarized Kerr black holes in extended scalar-tensor--Gauss-Bonnet gravity},
  journal = {Physical Review Letters},
  volume  = {123},
  number  = {1},
  pages   = {011101},
  year    = {2019},
  doi     = {10.1103/PhysRevLett.123.011101}
}

@article{GuoLi2020,
  author  = {Guo, Minyong and Li, Peng-Cheng},
  title   = {Innermost stable circular orbit and shadow of the 4D Einstein--Gauss--Bonnet black hole},
  journal = {European Physical Journal C},
  volume  = {80},
  number  = {6},
  pages   = {588},
  year    = {2020},
  doi     = {10.1140/epjc/s10052-020-8164-7}
}

@article{SilvaCoatesRamazanoglu2022,
  author  = {Silva, Hector O. and Coates, Andrew and Ramazano{\u{g}}lu, Fethi M. and Sotiriou, Thomas P.},
  title   = {The ghost of vector fields in compact stars},
  journal = {Physical Review D},
  volume  = {105},
  number  = {2},
  pages   = {024046},
  year    = {2022},
  doi     = {10.1103/PhysRevD.105.024046}
}

@article{Psaltis2008,
  author  = {Psaltis, Dimitrios},
  title   = {Probes and tests of strong-field gravity with observations in the electromagnetic spectrum},
  journal = {Living Reviews in Relativity},
  volume  = {11},
  number  = {1},
  pages   = {9},
  year    = {2008},
  doi     = {10.12942/lrr-2008-9}
}

@article{BardeenPressTeukolsky1972,
  author  = {Bardeen, James M. and Press, William H. and Teukolsky, Saul A.},
  title   = {Rotating black holes: Locally nonrotating frames, energy extraction, and scalar synchrotron radiation},
  journal = {The Astrophysical Journal},
  volume  = {178},
  pages   = {347},
  year    = {1972},
  doi     = {10.1086/151796}
}

@article{Bozza2002,
  author  = {Bozza, Valerio},
  title   = {Gravitational lensing in the strong field limit},
  journal = {Physical Review D},
  volume  = {66},
  pages   = {103001},
  year    = {2002},
  doi     = {10.1103/PhysRevD.66.103001},
  eprint  = {gr-qc/0208075},
  archivePrefix = {arXiv}
}

@article{CardosoEtAl2009,
  author  = {Cardoso, Vitor and Miranda, Alex S. and Berti, Emanuele and Witek, Helvi and Zanchin, Vilson T.},
  title   = {Geodesic stability, Lyapunov exponents and quasinormal modes},
  journal = {Physical Review D},
  volume  = {79},
  pages   = {064016},
  year    = {2009},
  doi     = {10.1103/PhysRevD.79.064016},
  eprint  = {0812.1806},
  archivePrefix = {arXiv},
  primaryClass  = {hep-th}
}

@article{Ramazanoglu2017,
  author  = {Ramazano{\u{g}}lu, Fethi M.},
  title   = {Spontaneous growth of vector fields in gravity},
  journal = {Physical Review D},
  volume  = {96},
  pages   = {064009},
  year    = {2017},
  doi     = {10.1103/PhysRevD.96.064009},
  eprint  = {1706.01056},
  archivePrefix = {arXiv},
  primaryClass  = {gr-qc}
}

@article{PizzutiPombo2024,
  author  = {Pizzuti, Lorenzo and Pombo, Alexandre M.},
  title   = {The spooky ghost of vectorization},
  journal = {Physics of the Dark Universe},
  volume  = {43},
  pages   = {101427},
  year    = {2024},
  doi     = {10.1016/j.dark.2024.101427},
  eprint  = {2310.18399},
  archivePrefix = {arXiv},
  primaryClass  = {gr-qc}
}

@misc{ChiangGarciaSaenzSang2026,
  author  = {Chiang, Hsu-Wen and Garcia-Saenz, Sebastian and Sang, Aofei},
  title   = {No-go theorem for spontaneous vectorization},
  year    = {2026},
  eprint  = {2605.13920},
  archivePrefix = {arXiv},
  primaryClass  = {gr-qc}
}

@misc{KluzniakAbramowicz2002,
  author  = {Klu{\'z}niak, W. and Abramowicz, M. A.},
  title   = {Parametric epicyclic resonance in black hole disks: QPOs in micro-quasars},
  year    = {2002},
  eprint  = {astro-ph/0203314},
  archivePrefix = {arXiv}
}

@article{JohannsenPsaltis2011,
  author={Johannsen, Tim and Psaltis, Dimitrios},
  title={Metric for rapidly spinning black holes suitable for strong-field tests of the no-hair theorem},
  journal={Physical Review D}, volume={83}, pages={124015}, year={2011},
  doi={10.1103/PhysRevD.83.124015}
}

@article{RezzollaZhidenko2014,
  author={Rezzolla, Luciano and Zhidenko, Alexander},
  title={New parametrization for spherically symmetric black holes in metric theories of gravity},
  journal={Physical Review D}, volume={90}, pages={084009}, year={2014},
  doi={10.1103/PhysRevD.90.084009}
}

@article{Krawczynski2012,
  author={Krawczynski, Henric},
  title={Tests of general relativity in the strong-gravity regime based on X-ray spectropolarimetric observations of black holes in X-ray binaries},
  journal={The Astrophysical Journal}, volume={754}, pages={133}, year={2012},
  doi={10.1088/0004-637X/754/2/133}
}

@article{Krawczynski2018,
  author={Krawczynski, Henric},
  title={Difficulties of quantitative tests of the Kerr-hypothesis with X-ray observations of mass accreting black holes},
  journal={General Relativity and Gravitation}, volume={50}, pages={100}, year={2018},
  doi={10.1007/s10714-018-2419-8}
}

@article{AbdujabbarovRezzollaAhmedov2015,
  author={Abdujabbarov, Ahmadjon and Rezzolla, Luciano and Ahmedov, Bobomurat},
  title={A coordinate-independent characterization of a black hole shadow},
  journal={Monthly Notices of the Royal Astronomical Society}, volume={454}, pages={2423--2435}, year={2015},
  doi={10.1093/mnras/stv2079}
}

@article{MizunoEtAl2018,
  author={Mizuno, Yosuke and Younsi, Ziri and Fromm, Christian M. and others},
  title={The current ability to test theories of gravity with black hole shadows},
  journal={Nature Astronomy}, volume={2}, pages={585--590}, year={2018},
  doi={10.1038/s41550-018-0449-5}
}

@article{VolkelKokkotas2019,
  author={V{\"o}lkel, Sebastian H. and Kokkotas, Kostas D.},
  title={Scalar fields and parametrized spherically symmetric black holes: Can one hear the shape of space-time?},
  journal={Physical Review D}, volume={100}, pages={044026}, year={2019},
  doi={10.1103/PhysRevD.100.044026}
}

@article{Johannsen2013,
  author={Johannsen, Tim},
  title={A Metric for Testing the Nature of Black Holes},
  journal={Journal of Physics: Conference Series}, volume={410}, pages={012136}, year={2013},
  doi={10.1088/1742-6596/410/1/012136}
}

@article{BertiEtAl2015,
  author={Berti, Emanuele and Barausse, Enrico and Cardoso, Vitor and others},
  title={Testing general relativity with present and future astrophysical observations},
  journal={Classical and Quantum Gravity}, volume={32}, pages={243001}, year={2015},
  doi={10.1088/0264-9381/32/24/243001}
}

@article{Will2014,
  author={Will, Clifford M.},
  title={The Confrontation between General Relativity and Experiment},
  journal={Living Reviews in Relativity}, volume={17}, pages={4}, year={2014},
  doi={10.12942/lrr-2014-4}
}

@article{BlazquezSalcedoKleihausKunz2021,
  author={Bl{\'a}zquez-Salcedo, Jos{\'e} Luis and Kleihaus, Burkhard and Kunz, Jutta},
  title={Scalarized black holes},
  journal={Arabian Journal of Mathematics}, volume={11}, pages={17--30}, year={2022},
  doi={10.1007/s40065-021-00349-7}
}

@article{BlazquezSalcedoEtAl2020,
  author={Bl{\'a}zquez-Salcedo, Jos{\'e} Luis and Doneva, Daniela D. and Kahlen, Sarah and others},
  title={Polar quasinormal modes of the scalarized Einstein-Gauss-Bonnet black holes},
  journal={Physical Review D}, volume={102}, pages={024086}, year={2020},
  doi={10.1103/PhysRevD.102.024086}
}

@article{AntoniouEtAl2022,
  author={Antoniou, Georgios and Macedo, Caio F. B. and McManus, Ryan M. and others},
  title={Stable spontaneously-scalarized black holes in generalized scalar-tensor theories},
  journal={Physical Review D}, volume={106}, pages={024029}, year={2022},
  doi={10.1103/PhysRevD.106.024029}
}

@article{MinamitsujiMukohyamaTsujikawa2024,
  author={Minamitsuji, Masato and Mukohyama, Shinji and Tsujikawa, Shinji},
  title={Angular and radial stabilities of spontaneously scalarized black holes in the presence of scalar-Gauss-Bonnet couplings},
  journal={Physical Review D}, volume={109}, pages={104057}, year={2024},
  doi={10.1103/PhysRevD.109.104057}
}

@article{KleihausKunzUtermohlen2023,
  author={Kleihaus, Burkhard and Kunz, Jutta and Uterm{\"o}hlen, Tim},
  title={Quadrupole instability of static scalarized black holes},
  journal={Physical Review D}, volume={107}, pages={L081501}, year={2023},
  doi={10.1103/PhysRevD.107.L081501}
}

@article{DonevaStaykovYazadjiev2019,
  author={Doneva, Daniela D. and Staykov, Kalin V. and Yazadjiev, Stoytcho S.},
  title={Gauss-Bonnet black holes with a massive scalar field},
  journal={Physical Review D}, volume={99}, pages={104045}, year={2019},
  doi={10.1103/PhysRevD.99.104045}
}

@article{DonevaCollodelKrugerEtAl2020,
  author={Doneva, Daniela D. and Collodel, Lucas G. and Kr{\"u}ger, Christian J. and Yazadjiev, Stoytcho S.},
  title={Spin-induced scalarization of Kerr black holes with a massive scalar field},
  journal={European Physical Journal C}, volume={80}, pages={1205}, year={2020},
  doi={10.1140/epjc/s10052-020-08765-3}
}

@article{OliveiraPombo2021,
  author={Oliveira, Jo{\~a}o M. S. and Pombo, Alexandre M.},
  title={Spontaneous vectorization of electrically charged black holes},
  journal={Physical Review D}, volume={103}, pages={044004}, year={2021},
  doi={10.1103/PhysRevD.103.044004}
}

@article{Matsumoto2023,
  author={Matsumoto, S.},
  title={No-go theorems for hairy black holes in scalar- or vector-tensor-Gauss--Bonnet theory},
  journal={Classical and Quantum Gravity}, volume={40}, pages={175011}, year={2023},
  doi={10.1088/1361-6382/ace94e}
}

@article{BartonKieferKleihaus2022,
  author={Barton, Simon and Kiefer, Claus and Kleihaus, Burkhard},
  title={Symmetric wormholes in Einstein-vector--Gauss--Bonnet theory},
  journal={European Physical Journal C}, volume={82}, pages={802}, year={2022},
  doi={10.1140/epjc/s10052-022-10761-8}
}

@article{RamazanogluUnluturk2019,
  author={Ramazano{\u{g}}lu, Fethi M. and {\"U}nl{\"u}t{\"u}rk, K{\i}van{\c{c}} {\.I}.},
  title={Generalized disformal coupling leads to spontaneous tensorization},
  journal={Physical Review D}, volume={100}, pages={084026}, year={2019},
  doi={10.1103/PhysRevD.100.084026}
}

@article{BrihayeHartmannKleihaus2022,
  author={Brihaye, Yves and Hartmann, Betti and Kleihaus, Burkhard and Kunz, Jutta},
  title={Horndeski-Proca stars with vector hair},
  journal={Physical Review D}, volume={105}, pages={044050}, year={2022},
  doi={10.1103/PhysRevD.105.044050}
}

@article{Fan2016,
  author={Fan, Zhong-Ying},
  title={Black holes with vector hair},
  journal={Journal of High Energy Physics}, volume={2016}, number={9}, pages={039}, year={2016},
  doi={10.1007/JHEP09(2016)039}
}

@article{Fan2018,
  author={Fan, Zhong-Ying},
  title={Black holes in vector-tensor theories and their thermodynamics},
  journal={European Physical Journal C}, volume={78}, pages={65}, year={2018},
  doi={10.1140/epjc/s10052-018-5540-7}
}

@article{BabichevCharmousisHassaine2017,
  author={Babichev, Eugeny and Charmousis, Christos and Hassaine, Mokhtar},
  title={Black holes and solitons in an extended Proca theory},
  journal={Journal of High Energy Physics}, volume={2017}, number={5}, pages={114}, year={2017},
  doi={10.1007/JHEP05(2017)114}
}

@article{HeisenbergTsujikawa2018,
  author={Heisenberg, Lavinia and Tsujikawa, Shinji},
  title={Hairy black hole solutions in U(1) gauge-invariant scalar--vector--tensor theories},
  journal={Physics Letters B}, volume={780}, pages={638--646}, year={2018},
  doi={10.1016/j.physletb.2018.03.059}
}

@article{HeisenbergKaseTsujikawa2018,
  author={Heisenberg, Lavinia and Kase, Ryotaro and Tsujikawa, Shinji},
  title={Odd-parity stability of hairy black holes in U(1) gauge-invariant scalar-vector-tensor theories},
  journal={Physical Review D}, volume={97}, pages={124043}, year={2018},
  doi={10.1103/PhysRevD.97.124043}
}

@article{HerdeiroRaduRunarsson2016,
  author={Herdeiro, Carlos A. R. and Radu, Eugen and R{\'u}narsson, Helgi Freyr},
  title={Kerr black holes with Proca hair},
  journal={Classical and Quantum Gravity}, volume={33}, pages={154001}, year={2016},
  doi={10.1088/0264-9381/33/15/154001}
}

@article{RahmanSen2019,
  author={Rahman, Mostafizur and Sen, Anjan A.},
  title={Astrophysical signatures of black holes in generalized Proca theories},
  journal={Physical Review D}, volume={99}, pages={024052}, year={2019},
  doi={10.1103/PhysRevD.99.024052}
}

@article{ZhouBambiHerdeiro2017,
  author={Zhou, Menglei and Bambi, Cosimo and Herdeiro, Carlos A. R. and Radu, Eugen},
  title={Iron K-alpha line of Kerr black holes with Proca hair},
  journal={Physical Review D}, volume={95}, pages={104035}, year={2017},
  doi={10.1103/PhysRevD.95.104035}
}

@article{ChannuieMomeni2018,
  author={Channuie, Phongpichit and Momeni, Davood},
  title={On the scalar-vector-tensor gravity: Black hole, thermodynamics and geometrothermodynamics},
  journal={Physics Letters B}, volume={785}, pages={309--314}, year={2018},
  doi={10.1016/j.physletb.2018.08.052}
}

@article{StefanovYazadjiev2010,
  author={Stefanov, Ivan Zh. and Yazadjiev, Stoytcho S.},
  title={Connection between Black-Hole Quasinormal Modes and Lensing in the Strong Deflection Limit},
  journal={Physical Review Letters}, volume={104}, pages={251103}, year={2010},
  doi={10.1103/PhysRevLett.104.251103}
}

@article{GalloVillanueva2015,
  author={Gallo, Emanuel and Villanueva, J. R.},
  title={Photon spheres in Einstein and Einstein-Gauss-Bonnet theories and circular null geodesics in axially-symmetric spacetimes},
  journal={Physical Review D}, volume={92}, pages={064048}, year={2015},
  doi={10.1103/PhysRevD.92.064048}
}

@article{KogaHarada2019,
  author={Koga, Yasutaka and Harada, Tomohiro},
  title={Stability of null orbits on photon spheres and photon surfaces},
  journal={Physical Review D}, volume={100}, pages={064040}, year={2019},
  doi={10.1103/PhysRevD.100.064040}
}

@article{GuoWangWu2022,
  author={Guo, Guangzhou and Wang, Peng and Wu, Houwen and Yang, Haitang},
  title={Quasinormal modes of black holes with multiple photon spheres},
  journal={Journal of High Energy Physics}, volume={2022}, number={6}, pages={060}, year={2022},
  doi={10.1007/JHEP06(2022)060}
}

@article{Tsupko2022,
  author={Tsupko, Oleg Yu.},
  title={Shape of higher-order images of equatorial emission rings around a Schwarzschild black hole: Analytical description with polar curves},
  journal={Physical Review D}, volume={106}, pages={064033}, year={2022},
  doi={10.1103/PhysRevD.106.064033}
}

@article{PerlickTsupko2022,
  author={Perlick, Volker and Tsupko, Oleg Yu.},
  title={Calculating black hole shadows: Review of analytical studies},
  journal={Physics Reports}, volume={947}, pages={1--39}, year={2022},
  doi={10.1016/j.physrep.2021.10.004}
}

@article{EHTM872019,
  author={{Event Horizon Telescope Collaboration}},
  title={First M87 Event Horizon Telescope Results. I. The Shadow of the Supermassive Black Hole},
  journal={The Astrophysical Journal Letters}, volume={875}, pages={L1}, year={2019},
  doi={10.3847/2041-8213/ab0ec7}
}

@article{EHTSgrA2022,
  author={{Event Horizon Telescope Collaboration}},
  title={First Sagittarius A* Event Horizon Telescope Results. I. The Shadow of the Supermassive Black Hole in the Center of the Milky Way},
  journal={The Astrophysical Journal Letters}, volume={930}, pages={L12}, year={2022},
  doi={10.3847/2041-8213/ac6674}
}

@article{Synge1966,
  author={Synge, J. L.},
  title={The Escape of Photons from Gravitationally Intense Stars},
  journal={Monthly Notices of the Royal Astronomical Society}, volume={131}, pages={463--466}, year={1966},
  doi={10.1093/mnras/131.3.463}
}

@article{Darwin1959,
  author={Darwin, Charles Galton},
  title={The gravity field of a particle},
  journal={Proceedings of the Royal Society of London A}, volume={249}, pages={180--194}, year={1959},
  doi={10.1098/rspa.1959.0015}
}

@article{Perlick2004,
  author={Perlick, Volker},
  title={Gravitational Lensing from a Spacetime Perspective},
  journal={Living Reviews in Relativity}, volume={7}, pages={9}, year={2004},
  doi={10.12942/lrr-2004-9}
}

@article{CunhaHerdeiroRadu2017,
  author={Cunha, Pedro V. P. and Herdeiro, Carlos A. R. and Radu, Eugen},
  title={Fundamental photon orbits: Black hole shadows and spacetime instabilities},
  journal={Physical Review D}, volume={96}, pages={024039}, year={2017},
  doi={10.1103/PhysRevD.96.024039}
}

@article{ClaudelVirbhadraEllis2001,
  author={Claudel, Clarissa-Marie and Virbhadra, K. S. and Ellis, G. F. R.},
  title={The geometry of photon surfaces},
  journal={Journal of Mathematical Physics}, volume={42}, pages={818--838}, year={2001},
  doi={10.1063/1.1308507}
}

@article{HiokiMaeda2009,
  author={Hioki, Kenta and Maeda, Kei-ichi},
  title={Measurement of the Kerr spin parameter by observation of a compact object's shadow},
  journal={Physical Review D}, volume={80}, pages={024042}, year={2009},
  doi={10.1103/PhysRevD.80.024042}
}

@article{PageThorne1974,
  author={Page, Don N. and Thorne, Kip S.},
  title={Disk-Accretion onto a Black Hole. Time-Averaged Structure of Accretion Disk},
  journal={The Astrophysical Journal}, volume={191}, pages={499--506}, year={1974},
  doi={10.1086/152990}
}

@article{PennaEtAl2010,
  author={Penna, Robert F. and McKinney, Jonathan C. and Narayan, Ramesh and others},
  title={Simulations of magnetized discs around black holes: effects of black hole spin, disc thickness and magnetic field geometry},
  journal={Monthly Notices of the Royal Astronomical Society}, volume={408}, pages={752--782}, year={2010},
  doi={10.1111/j.1365-2966.2010.17170.x}
}

@article{AbramowiczKluzniak2001,
  author={Abramowicz, Marek A. and Klu{\'z}niak, W{\l}odek},
  title={A precise determination of black hole spin in GRO J1655-40},
  journal={Astronomy and Astrophysics}, volume={374}, pages={L19--L20}, year={2001},
  doi={10.1051/0004-6361:20010791}
}

@article{TorokEtAl2011,
  author={T{\"o}r{\"o}k, Gabriel and Kotrlov{\'a}, Andrea and {\v S}r{\'a}mkov{\'a}, Eva and others},
  title={Confronting the models of 3:2 quasiperiodic oscillations with the rapid spin of the microquasar GRS 1915+105},
  journal={Astronomy and Astrophysics}, volume={531}, pages={A59}, year={2011},
  doi={10.1051/0004-6361/201015549}
}

@article{SramkovaEtAl2015,
  author={{\v S}r{\'a}mkov{\'a}, Eva and T{\"o}r{\"o}k, Gabriel and Kotrlov{\'a}, Andrea and others},
  title={Black hole spin inferred from 3:2 epicyclic resonance model of high-frequency quasi-periodic oscillations},
  journal={Astronomy and Astrophysics}, volume={578}, pages={A90}, year={2015},
  doi={10.1051/0004-6361/201425241}
}

@article{StuchlikKolos2015,
  author={Stuchl{\'i}k, Zden{\v e}k and Kolo{\v s}, Martin},
  title={Mass of intermediate black hole in the source M82 X-1 restricted by models of twin high-frequency quasi-periodic oscillations},
  journal={Monthly Notices of the Royal Astronomical Society}, volume={451}, pages={2575--2588}, year={2015},
  doi={10.1093/mnras/stv1120}
}

@article{StuchlikKolos2016,
  author={Stuchl{\'i}k, Zden{\v e}k and Kolo{\v s}, Martin},
  title={Models of quasi-periodic oscillations related to mass and spin of the GRO J1655-40 black hole},
  journal={Astronomy and Astrophysics}, volume={586}, pages={A130}, year={2016},
  doi={10.1051/0004-6361/201526095}
}

@article{Stefanov2014,
  author={Stefanov, Ivan Zh.},
  title={Confronting models for the high-frequency QPOs with Lense--Thirring precession},
  journal={Monthly Notices of the Royal Astronomical Society}, volume={444}, pages={2178--2185}, year={2014},
  doi={10.1093/mnras/stu1602}
}

@article{RemillardMcClintock2006,
  author={Remillard, Ronald A. and McClintock, Jeffrey E.},
  title={X-Ray Properties of Black-Hole Binaries},
  journal={Annual Review of Astronomy and Astrophysics}, volume={44}, pages={49--92}, year={2006},
  doi={10.1146/annurev.astro.44.051905.092532}
}

@article{MottaEtAl2014,
  author={Motta, S. and Belloni, T. and Stella, L. and Mu{\~n}oz-Darias, T. and Fender, R.},
  title={Precise mass and spin measurements for a stellar-mass black hole through X-ray timing: the case of GRO J1655-40},
  journal={Monthly Notices of the Royal Astronomical Society}, volume={437}, pages={2554--2565}, year={2014},
  doi={10.1093/mnras/stt2068}
}

@article{StellaVietri1998,
  author={Stella, Luigi and Vietri, Mario},
  title={Lense-Thirring Precession and Quasi-periodic Oscillations in Low-Mass X-Ray Binaries},
  journal={The Astrophysical Journal Letters}, volume={492}, pages={L59--L62}, year={1998},
  doi={10.1086/311075}
}

@article{StellaVietriMorsink1999,
  author={Stella, Luigi and Vietri, Mario and Morsink, Sharon M.},
  title={Correlations in the Quasi-periodic Oscillation Frequencies of Low-Mass X-Ray Binaries and the Relativistic Precession Model},
  journal={The Astrophysical Journal Letters}, volume={524}, pages={L63--L66}, year={1999},
  doi={10.1086/312291}
}

@article{AbramowiczFragile2013,
  author={Abramowicz, Marek A. and Fragile, P. Chris},
  title={Foundations of Black Hole Accretion Disk Theory},
  journal={Living Reviews in Relativity}, volume={16}, pages={1}, year={2013},
  doi={10.12942/lrr-2013-1}
}

\end{document}